\documentclass[a4paper,fleqn]{cas-dc}

\usepackage[numbers]{natbib}

\usepackage{multirow}

\def\tsc#1{\csdef{#1}{\textsc{\lowercase{#1}}\xspace}}
\tsc{WGM}
\tsc{QE}

\begin{document}
\let\WriteBookmarks\relax
\def\floatpagepagefraction{1}
\def\textpagefraction{.001}

\shorttitle{Development of wide range photon detection system for muonic X-ray spectroscopy}

\shortauthors{Mizuno \textit{et al.}}

\title[mode=title]{Development of wide range photon detection system for muonic X-ray spectroscopy}  



%

\author[1]{R. Mizuno}

\cormark[1] 

\fnmark[1]

\ead{mizuno@nex.phys.s.u-tokyo.ac.jp}



\affiliation[1]{organization={Department of Physics, Faculty of Science, The University of Tokyo},
            addressline={Hongo 7-3-1}, 
            city={Bunkyo},
            postcode={113-0033}, 
            state={Tokyo},
            country={Japan}}

\author[1,2]{M.~Niikura}
\author[3,1,4]{T.Y.~Saito}
\author[2]{T.~Matsuzaki}
\author[1,2]{H.~Sakurai}
\author[5]{A.~Amato}
\author[6]{S.~Asari}
\author[5]{S.~Biswas}
\author[7]{I.~Chiu}
\author[5]{L.~Gerchow}
\author[5]{Z.~Guguchia}
\author[5]{G.~Janka}
\author[8]{K.~Ninomiya}
\author[9]{N.~Ritjoho}
\author[10]{A.~Sato}
\author[11,5]{K.~von~Schoeler}
\author[12]{D.~Tomono}
\author[13]{K.~Terada}
\author[5]{C.~Wang}

\affiliation[2]{organization={RIKEN Nishina Center},
            addressline={Hirosawa 2-1}, 
            city={Wako},
            postcode={351-0198}, 
            state={Saitama},
            country={Japan}}
\affiliation[3]{organization={Atomic, Molecular, and Optical Physics Laboratory, RIKEN},
            addressline={Hirosawa 2-1}, 
            city={Wako},
            postcode={351-0198}, 
            state={Saitama},
            country={Japan}}
\affiliation[4]{organization={Center for Nuclear Study, Faculty of Science, The University of Tokyo},
            addressline={Hirosawa 2-1}, 
            city={Wako},
            postcode={351-0198}, 
            state={Saitama},
            country={Japan}}
\affiliation[5]{organization={Paul Scherrer Institut},
            addressline={Forschungsstrasse 111}, 
            city={Villigen},
            postcode={5232}, 
            country={Switzerland}}
\affiliation[6]{organization={Department of Chemistry, Graduate School of Science, Osaka Unviersity},
            addressline={Machikaneyamacho 1-1}, 
            city={Toyonaka},
            postcode={560-0043}, 
            state={Osaka},
            country={Japan}}
\affiliation[7]{organization={Japan Atomic Energy Agency (JAEA)},
            addressline={Shirakata 2-4},
            city={Nakagun Tokaimura},
            postcode={319-1195},
            state={Ibaraki},
            country={Japan}}
\affiliation[8]{organization={Institute for Radiation Science, Osaka Unviersity},
            addressline={Machikaneyama-cho 1-1}, 
            city={Toyonaka},
            postcode={560-0043}, 
            state={Osaka},
            country={Japan}}
\affiliation[9]{organization={Suranaree University of Technology},
            addressline={111, Maha Witthayalai Rd, Suranari, Mueang Nakhon Ratchasima District, Nakhon Ratchasima},
            postcode={30000},
            country={Thailand}}
\affiliation[10]{organization={Department of Physics, Graduate School of Science, Osaka Unviersity},
            addressline={Machikaneyamacho 1-1}, 
            city={Toyonaka},
            postcode={560-0043}, 
            state={Osaka},
            country={Japan}}
\affiliation[11]{organization={Institute for Particle Physics and Astrophysics, ETH Zurich},
            addressline={Otto-Stern-Weg 5},
            city={Zurich},
            postcode={8093},
            country={Switzerland}}
\affiliation[12]{organization={Research Center for Nuclear Physics, Osaka Unviersity},
            addressline={Mihogaoka 10-1}, 
            city={Ibaraki},
            postcode={567-0047}, 
            state={Osaka},
            country={Japan}}
\affiliation[13]{organization={Department of Earth and Space Science, Graduate School of Science, Osaka Unviersity},
            addressline={Machikaneyama-cho 1-1}, 
            city={Toyonaka},
            postcode={560-0043}, 
            state={Osaka},
            country={Japan}}


\begin{abstract}
We have developed a photon detection system for muonic X-ray spectroscopy.
The detector system consists of high-purity germanium detectors with BGO Compton suppressors.
The signals from the detectors are readout with a digital acquisition system.
The absolute energy accuracy, energy and timing resolutions, photo-peak efficiency, the performance of the Compton suppressor, and high count rate durability are studied with standard $\gamma$-ray sources and in-beam experiment using $^{27}\mathrm{Al}(p, \gamma){}^{28}\mathrm{Si}$ resonance reaction.
The detection system was demonstrated at Paul Scherrer Institute.
A calibration method for a photon detector at a muon facility using muonic X-rays of $^{197}$Au and $^{209}$Bi is proposed. 
\end{abstract}



\begin{keywords}
HPGe detector\sep muonic X-ray spectroscopy \sep Compton suppressor
\end{keywords}

\maketitle

\section{Introduction}

A muonic atom is a bound system consisting of a negative muon and a nucleus surrounded by electrons, formed 
when a negative muon stops in the matter. 
As the lifetime of the muon is sufficiently longer than the time scale of the atomic transition, X-rays originating from muonic transitions are emitted. 
As the muon mass is approximately 207 times heavier than the electron mass, muonic X-rays have higher energy than electric X-rays. 
Measurements of muonic X-rays are utilized in several fields of the natural and social sciences, including nuclear, atomic, and particle physics, chemistry, earth and planetary sciences, archaeology, and industrial applications.

Since muonic X-rays have characteristic energies for each element, they are used for nondestructive elemental analysis, the so-called ``muon induced X-ray emission (MIXE).'' 
MIXE is a nondestructive, three-dimensionally position selective, and simultaneous multi-element analysis, which can analyze several elements from lithium to uranium in bulk samples. 
MIXE has been utilized for several applications, such as the analysis of meteorites~\cite{Terada2014-cw, Terada2017-kb, Hofmann2023-kz, Chiu2023-mg}, asteroid samples~\cite{Nakamura2023-ge, Ninomiya2023}, lithium batteries~\cite{Umegaki2020-xi}, and archaeological samples~\cite{Ninomiya2015-jh, Hampshire2019-tc, Shimada-Takaura2021-ae, Biswas2023-xz}. 


Muonic X-rays are also used as a method to measure the nuclear charge radius and distributions~\cite{Fitch1953-et, Fricke1995-cm, Angeli2013-to, Saito2022-wl, Antognini2020-zf}. 
Because the muon has a large mass and the atomic radius of the muonic atom is considerably smaller than that of an ordinary electron, the binding energy of the muonic atom is highly sensitive to the charge distribution of the nuclei. 
To measure the charge radius using muonic X-ray spectroscopy, the accuracy of the measured energy is essential.
The energies of the Lyman series of the muonic transition range from a few tens keV to 6 MeV depending on the atomic number.
To conduct charge radius measurements using muonic X-ray spectroscopy, a highly accurate X-ray measurement system that can apply to a wide energy range is required. 

Therefore, in this study, we developed a wide-energy-range photon detection system for MIXE and nuclear charge radius measurements. 
The detection system is designed for performing muonic X-ray spectroscopy at continuous muon beam facilities such as the Research Center for Nuclear Physics (RCNP), Paul Scherrer Institute (PSI), and Canada's Particle Accelerator Centre (TRIUMF).
High-purity germanium (Ge) detectors are used for high energy resolution X-ray spectroscopy. 
Compton suppressors using bismuth germanium oxide (BGO) scintillators were adopted to reduce the background noise because detecting muonic X-rays from light elements is limited by the signal-to-noise (S/N) ratio in low-energy regions.
The data acquisition was constructed using a waveform digitizer in the system to accommodate for high count-rate measurements at continuous muon beam facilities. 
A Monte Carlo simulation was performed to design the experimental setup and optimize the detector geometry. 
The $^{27}\mathrm{Al}(p,\gamma){}^{28}\mathrm{Si}$ resonance reaction was used to evaluate the response of the photon detector for high-energy muonic X-rays emitted from heavy elements~\cite{Mizuno2023-px}. 
As the calibration method using the resonance reaction is not applicable at muon facilities, a new calibration method using muonic X-rays from $^{197}$Au and $^{209}$Bi as energy and intensity references was proposed. 

The remainder of this paper is organized as follows: 
The design concept of the detection system is described in Sect.~\ref{sec:Methods_1}. 
The methods used for evaluating the detector system are presented in Sect.~\ref{sec:Methods_2} and Sect.~\ref{sec:Methods_3}. 
The performance evaluation of the detector in terms of linearity, energy resolution, timing resolution, photopeak efficiency, and Compton suppression is presented in Sect.~\ref{sec:linearity}--\ref{sec:compton}.
The measurement results of muonic X-rays from a meteorite, $^{197}$Au, $^{208}$Pb, $^{209}$Bi are discussed in Sect.~\ref{sec:MIXE}. 
Finally, the conclusions of the present study are presented in Sect.~\ref{sec:summary}.

\section{Design of the detection system} \label{sec:Methods_1}

For muonic X-ray spectroscopy, a large dynamic range and high energy resolution are required.
Ge detector arrays~\cite{Gerchow2023-pp, Tampo2023-qf, Hillier2016-xa} are generally used in the muonic X-ray detection; however, in recent years, cryogenic quantum sensors ~\cite{Ullom2015-yb} and CdTe detectors~\cite{Chiu2022-me} are also used. 
Ge detectors show a typical energy dynamic range from 1 keV to 10 MeV with an energy resolution of $\Delta E/E\sim0.1\%$. In contrast, cryogenic quantum sensors show approximately 1--10 keV dynamic range and 0.03\% energy resolution, and CdTe detectors show approximately 10--150 keV dynamic range and 1.2\% energy resolution.
The energies of a muonic $K_\alpha$ X-ray of low atomic number ($Z$) elements, such as carbon, nitrogen and oxygen, are below 150 keV, whereas that of large $Z$ elements, such as the actinides, are above 6 MeV. 
To detect the $K_\alpha$ X-rays for high $Z$ elements for nuclear charge radii measurements, high energy resolution in a wide energy region is required; thus, we adopted the Ge detector as a photon detector.

As achieving a wide dynamic range and high energy resolution in a low-energy photon with a single type of Ge detector is difficult, various types of Ge detectors are combined to form a detector array.
A planar-type Ge detector with a thin entrance window is used for the low-energy region, and a large-volume coaxial Ge detector with sufficient efficiency is used for the high-energy region. 
For this reason, three types of Ge detectors, namely, Canberra BE2820, GX5019, and GC3018, were selected in this study.
Table~\ref{tab:Ge-character} shows the characteristics of each detector.
BE2820 is a broad-energy Ge detector consisting of a planar-shaped crystal with point contact electrode having the best energy resolution. 
BE2820 covers the energy range from 3 keV to 3 MeV owing to its thin carbon composite window and thin front end of the crystal. 
GX5019 is a p-type extended-range coaxial Ge detector with a large crystal volume and a thin-window contact on the front surface and shows relatively large efficiency ranging from 3 keV to more than 10 MeV.
GC3018 is a standard p-type coaxial detector 
and is used as a reference detector to BE2820 and GX5019 because GC3018 shows Ge detector's typical performances.
The relative efficiencies of BE2820, GX5019, and GC3018 are 13, 50, and 30\%, respectively. 
All the preamplifiers built-in the Ge detectors are the resistive feedback type.
The preamplifiers of BE2820 and GX5019 are intelligent preamplifier (iPA)-SL10, which can change the output gain in times 1, 2, 5, and 10 through a control system via a USB interface. 
The preamplifier of GC3018 can also change the output gain in times 1 and 5.
Notably 
the end-cap diameters of BE2820 and GX5019 are modified to 76 mm, which is smaller than their original catalog sizes and is the same diameter as that of GC3018 such that all detectors accommodate the same geometry of the BGO Compton suppressors explained below. 
In the present study, two BE2820, one GX5019, and two GC3018 were investigated.

\begin{table}
  \caption{The characteristics of Ge detectors. All the detectors are manufactured by Canberra Inc.}
  \label{tab:Ge-character}
 \begin{tabular*}{\tblwidth}{@{} llll@{} }
  \toprule
  Ge model         & GC3018      & GX5019     & BE2820 \\
  \midrule
  Crystal          & coaxial     & coaxial    & point contact planar \\
  Type             & P-type      & P-type     & N-type \\
  PreAmp           & 2002CSI     & iPA-SL10   & iPA-SL10 \\
  Rel. efficiency  & 30\%        & 50\%       & 13\% \\ 
  Window           & 1.6 mm (Al) & 0.6 mm (C) & 0.6 mm (C) \\
  Serial number    & 9681, 9682  & 5536       & 13385, 13386\\
  \bottomrule
  \end{tabular*}
\end{table}

In the MIXE experiments, especially to obtain a low composite from low $Z$ materials, the signal-to-noise (S/N) ratio needs to be small. 
In some MIXE experiments for earth and planetary materials, the low $Z$ element is measured in the presence of higher Z materials contamination; 
for example, carbon, nitrogen, and oxygen are measured with the background of silicon and iron. 
In such cases, the lowest detectable composition of lower $Z$ materials is limited to 1 wt\% because the K$_\alpha$ peaks of lower $Z$ elements are on the large Compton backgrounds of muonic X-rays from higher $Z$ elements~\cite{Terada2014-cw}. 
Therefore, a system to reduce the background component and improve the S/N ratio in low energy regions is required to identify the low composite from low $Z$ elements. 
The Compton suppression technique helps to reduce the background in the low-energy region.
Furthermore, the Compton suppressor helps to avoid cross-talk events. 
Note that thinner Ge detectors are also effective in reducing the Compton backgrounds in low-energy regions owing to low efficiency for high-energy photons. However, the detection efficiency is also reduced with thinner Ge detectors, which have smaller detection areas. The Compton suppressors with BE2820, GX5019, and GC3018 are adopted in this study because they provide both the reasonable detection efficiency and the better S/N ratio simultaneously.

\begin{figure*}[width=0.85\textwidth, cols=4,pos=h]
  \centering
  \includegraphics[scale=0.45]{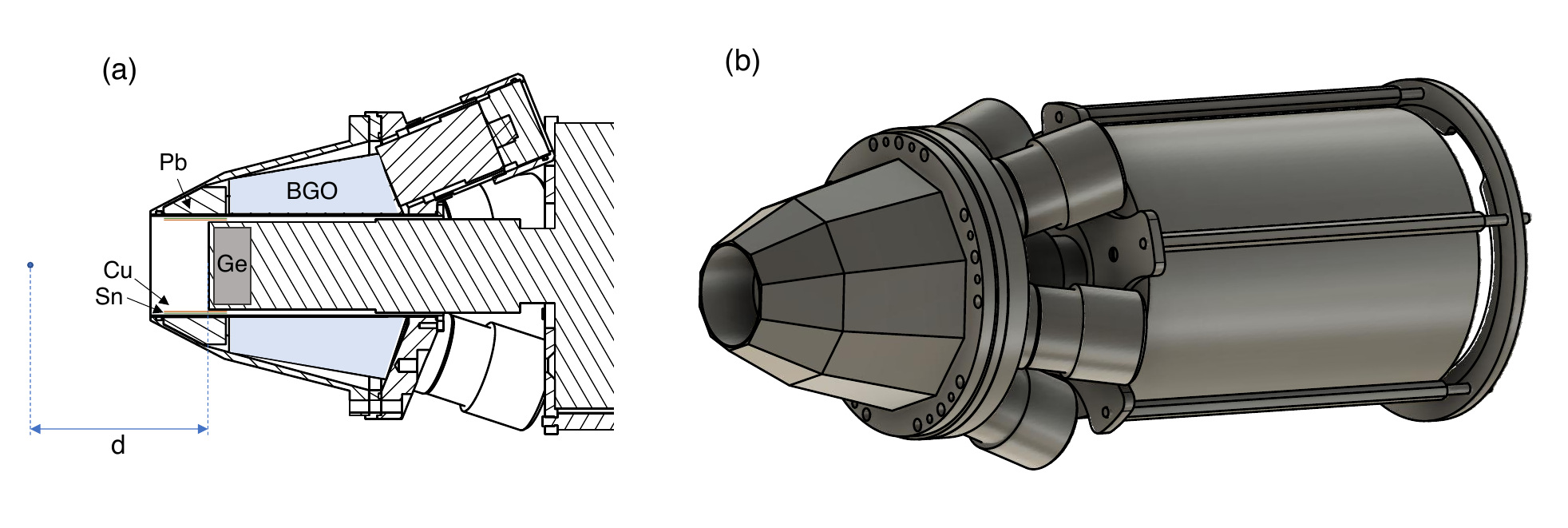}
  \caption{Cross-sectional view (a) and schematics (b) of the Ge detector with Compton suppressors. The point in figure~(a) shows the target or source position, and the Ge position ($d$) is defined as the distance from the target/source to the front side of the window of the Ge detector.}
  \label{fig:Ge_BGO}
\end{figure*}

BGO crystals surrounding the Ge detector were used for the Compton suppressor, as shown in Fig.~\ref{fig:Ge_BGO}. 
This Compton suppressor was originally developed for nuclear structure studies at the University of Tsukuba~\cite{BGO_tsukuba1, BGO_tsukuba2, BGO_tsukuba3}, and the detector geometry was modified to fit our Ge detectors in this study.
Five separated BGO crystals making a decagonal shape with a center hole of 80-mm in diameter were used, and 
the thickness of each crystal was approximately 22 mm and 44 mm for the front and back sides, respectively. 
The BGO crystals were placed 
such that they suppress forward scattered photons at the Ge crystal, which corresponds to the low-energy Compton region of the energy spectrum. 
A lead ring with 5-cm in length was placed in front of the BGO crystals 
to avoid the direct incidence of photons and particles from the target on the BGO crystals.
Cylindrical copper and tin plates were inserted on the inner surface of the lead ring to absorb electric X-rays originating from lead and tin with energies at 73, 75 keV, and 25, 28 keV, respectively.
The optical readout of the BGO scintillator was performed using photomultiplier tubes (PMT).
R6231 manufactured by Hamamatsu Photonics K. K. was used to achieve high sensitivity at 480 nm, which is the wavelength of the scintillation photon.
The Ge position ($d$) is defined as the distance from the target/source to the front side of the window of the Ge detector.

A data acquisition (DAQ) system using waveform digitizers was adapted because a high counting rate is expected in the muon irradiation experiment at Paul Scherrer Institute (PSI) owing to its high-intensity continuous beam.
The advantages of the DAQ system using digitizers are: fewer limitations in parameter tuning for data acquisition,
reduction in the number of circuit modules, and lower dead time because of fast processing.
The reduction in the number of modules enables lower cost and higher stability due to temperature changes.

All the signals from the detectors were acquired using 500-MS/s 14-bit waveform digitizers, CAEN V1730B~\cite{V1730B}.
V1730B consists of a 16-channel flash ADC and FPGA and can acquire digital waveforms, energy, and timing from input signals using FPGA firmware.
The dynamic input range of V1730B is selectable, 0.5 V and 2.0 V.
Two firmware on V1730B are used in the DAQ systems: digital pulse processing for pulse height analysis (DPP-PHA) 
and DPP for charge integration and pulse shape discriminator (DPP-PSD)~\cite{DPP-PHAandPSD}.
DPP-PHA deduces the pulse height of the input signals using the trapezoidal filter and the timing using the RC-CR$^2$ method.
DPP-PSD obtains the charge integration and the timing using the constant fraction discriminator (CFD) or the leading-edge timing (LET) method.
This DAQ system was controlled using CAEN Multi-Parameter Spectroscopy Software (CoMPASS), a DAQ software developed by CAEN S.p.A.~\cite{CoMPASS}.

\begin{figure}
  \centering
  \includegraphics[scale=0.5]{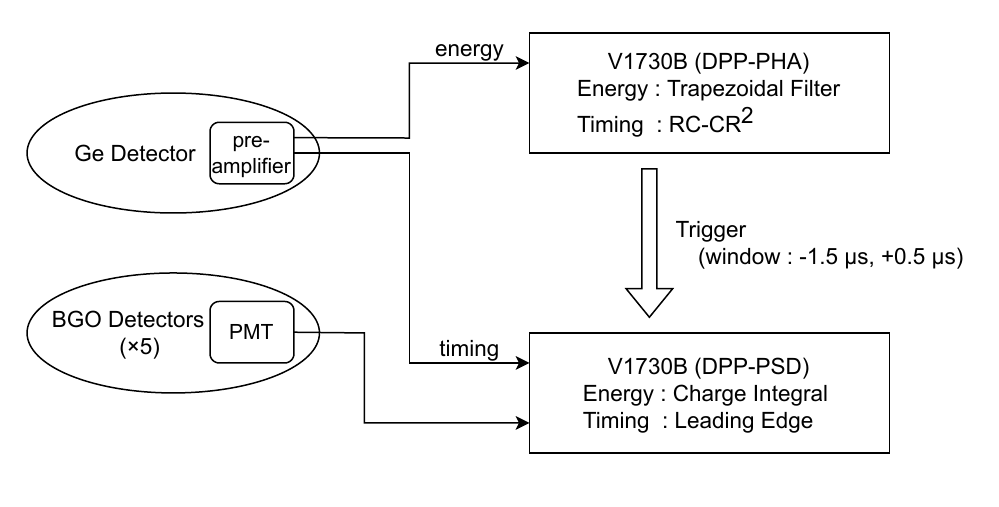}
   \caption{Schematics of the data acquisition system.
   The Ge energy was obtained using the DPP-PHA firmware, and the BGO energy and Ge timing were obtained using the DPP-PSD firmware.}
   \label{fig:DAQ}
\end{figure}

Figure~\ref{fig:DAQ} shows the schematics of the DAQ system using two V1730B boards.
The firmwares of the digitizer boards were DPP-PHA and DPP-PSD, respectively.
Two pre-amplifier signals from one Ge detector were put in the two digitizer boards:
one for energy information obtained using DPP-PHA and the other for timing information obtained using DPP-PSD.
This complexity is because the timing pick-off method of DPP-PHA firmware does not obtain sufficient timing resolution of the Ge detectors (see Sect.~\ref{sec:timing}). 
PMT signals from BGO scintillators were processed using DPP-PSD firmware to obtain energy and timing.
All signals from the Ge detectors were obtained and saved in the self-trigger mode, 
whereas BGO signals were obtained 
only in coincidence with the Ge detector. 
The coincidence time window between the Ge and BGO detectors was set to -1.5 to 0.5 $\mu$s corresponding to sufficient time to obtain all the coincident signals. 

\section{Performance of the detectors}
\label{sec:Performance}
The performance of the detectors was evaluated through offline measurements using standard $\gamma$-ray sources and an in-beam measurement using the $^{27}$Al(p,$\gamma$)$^{28}$Si resonance reaction prior to the demonstration of the detector system with muonic X-ray measurements at PSI.
The measurement setups for the offline and in-beam experiments are explained in Sect.~\ref{sec:Methods_2}.
A Monte Carlo simulation was performed to optimize the detector geometry and design muonic X-ray spectroscopy setups, as explained in Sect.~\ref{sec:Methods_3}. 
The evaluation of the performances of the Ge detectors as well as the energy accuracy, energy resolution, durability under the high count rate condition, timing resolution, photopeak efficiency, and Compton suppression results are presented in Sect.~\ref{sec:linearity}--\ref{sec:compton}, respectively.

\subsection{Measurements} \label{sec:Methods_2}
Measurements using standard $\gamma$-ray sources were performed 
to investigate the essential feature of Ge detectors and Compton suppressors used in the photon detection system using the waveform digitizer. 
The linearity, energy and timing resolutions, photo-peak detection efficiency, high count rate durability of the Ge detectors, and the performance of Compton suppression were evaluated using standard $\gamma$-ray sources of 
$^{152}$Eu, $^{60}$Co, $^{133}$Ba, and $^{137}$Cs in low-energy regions below 1.4 MeV.

An in-beam measurement using a proton-induced resonance reaction of $^{27}$Al at 992 keV, $^{27}\mathrm{Al}(p,\gamma){}^{28}\mathrm{Si}$, was conducted at the RIKEN tandem accelerator for evaluating the energy resolution and efficiency of the high-energy $\gamma$ rays for GX5019. 
The details of the experiment are presented in Ref.~\cite{Mizuno2023-px}.
The Ge detectors' photo-peak efficiencies and energy resolutions were deduced in a wide energy range from 1.5--10.8 MeV.

\subsection{GEANT4 simulation}\label{sec:Methods_3}
A Monte Carlo simulation using the GEANT4 toolkit~\cite{Agostinelli2003-eh} was performed to optimize the detector geometry and design the muonic X-ray spectroscopy setup.
In the present study, \texttt{GEANT4 v10.06.p03} was used, and \texttt{EmLivermorePhysics} was used as the physics model of electromagnetic interaction.
The reproducibility of the simulation was evaluated in terms of detection efficiency and Compton suppression, as presented in Sect.~\ref{sec:efficiency} and \ref{sec:compton}, respectively.

All the possible geometries were included in the simulation.
The geometries of the Ge crystals were taken from the specification sheets provided by the manufacturer.
The measured shape of the BGO crystals was implemented.
Other materials of the detector, such as the case of the detector, cryostat, entrance window, cold finger, liquid nitrogen dewar, BGO crystal holder, and lead ring were included in the simulation.
Furthermore, a target holder and other materials around the target/source were included in each situation. 
All these materials are important for reproducing the shape of the $\gamma$-ray spectra.
\subsection{Energy accuracy}\label{sec:linearity}
The energy accuracy of the detector system is critical for the measurement of nuclear charge radii as the accuracy of the muonic X-ray energy measurement is directly reflected in that of the nuclear charge radii.
For example, 0.2-keV accuracy on the X-ray energy corresponds to 0.66\% accuracy of the charge radius in $^{108}$Pd~\cite{Saito2022-wl} and 0.04\% in $^{208}$Pb~\cite{Bergem1988-nf}, respectively.
The energy accuracy is investigated in this section.

\begin{figure*}
  \centering
  \includegraphics[width=1.1\textwidth,scale=0.4]{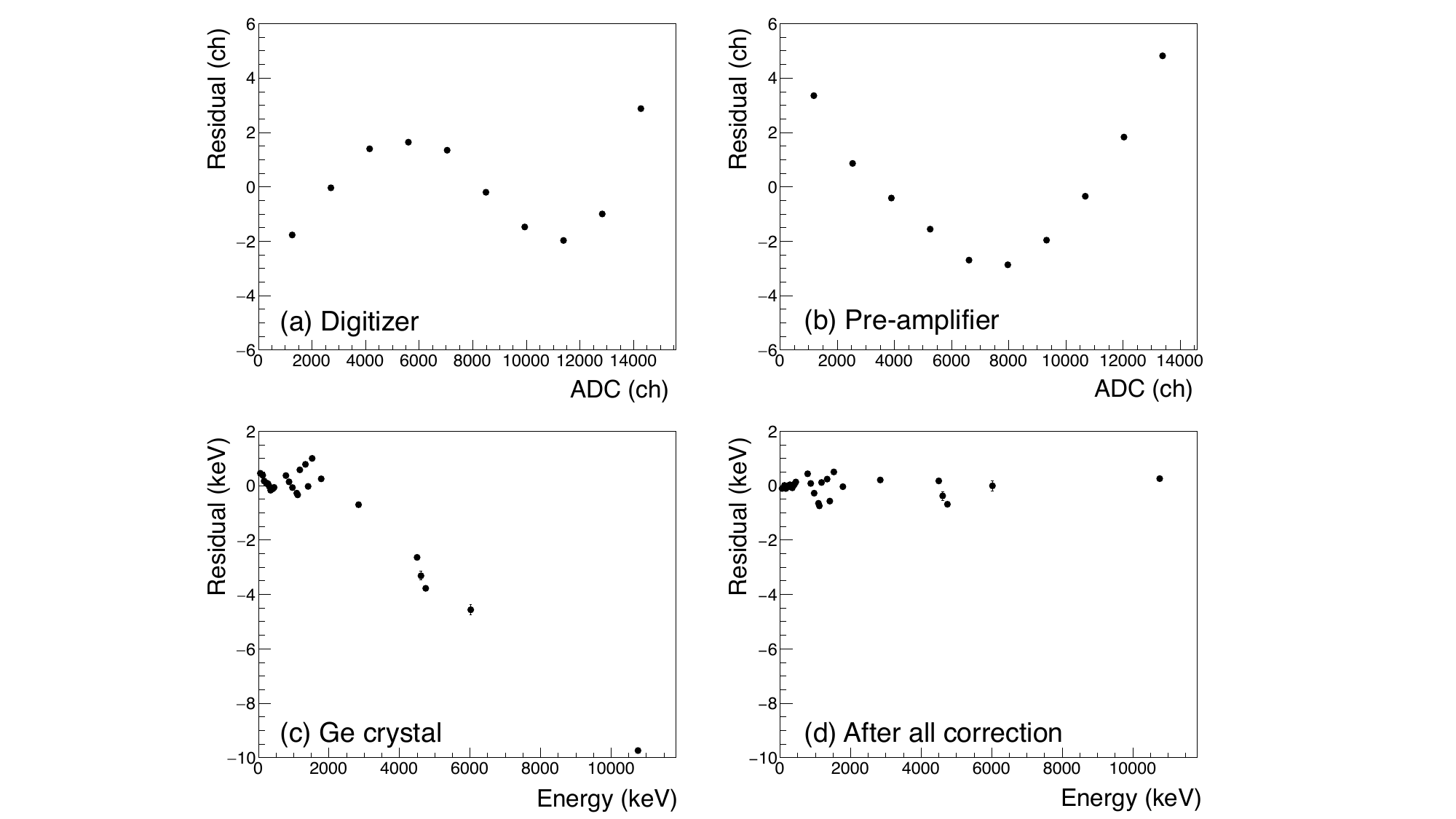}
  \caption{Residuals from the linear calibration function for (a) the digitizer (V1730B with a 2.0-V range), (b) the preamplifier of GX5019, (c) the Ge crystal of GX5019, and (d) after nonlinearity correction~\cite{Mizuno2023-px}.}
  \label{fig:linearity}
\end{figure*}

The statistical uncertainty is usually very small in the muonic X-ray spectroscopy and the energy accuracy is primarily limited by the systematic uncertainty in the energy calibration, namely, the nonlinearity of the detector system.
The linearities of the waveform digitizer, preamplifier, and Ge crystals were independently evaluated using a pulse generator (Ortec 448) and $\gamma$-ray measurements in the range of 50 keV to 10 MeV.
First, the nonlinearity of the digitizer was examined using the pulse generator.
The pulse signals were inputted into the digitizer with the DPP-PHA firmware, and the heights of the pulse generator were changed in 10 steps.
The nonlinearity of the digitizer was found in the residuals from the linear calibration function, as shown in Fig.~\ref{fig:linearity}(a).
The nonlinearity of the digitizer was corrected by a third-order polynomial function.
The constant term of the calibration function was fixed to zero because offset cancellation is included in digital pulse processing.
The residuals from the calibration curve became sufficiently smaller than the channel width of the digitizer.
Second, the nonlinearities of the pre-amplifiers of each Ge detector were evaluated by introducing a pulse generator signal into the test input of each preamplifier built in the Ge detectors. 
Figure~\ref{fig:linearity}(b) shows the residuals from the linear calibration function for the preamplifier of GX5019 after excluding the nonlinearity of the digitizer, which shows a higher-order trend.
Other detectors of GC3018 and BE2820 showed the same trends in the nonlinearity, and a third-order polynomial for all three detectors was used for the correction function to eliminate the nonlinearity.
Finally, the nonlinearity of the Ge crystals was measured with standard $\gamma$-ray sources.
For GX5019, the in-beam measurement of the $^{27}\mathrm{Al}(p,\gamma){}^{28}\mathrm{Si}$ reaction was also used to evaluate the nonlinearity of high-energy $\gamma$-rays.
Figure~\ref{fig:linearity}(c) shows the residuals from the linear calibration function for the Ge crystal of GX5019 after excluding the nonlinearity of the digitizer and preamplifier~\cite{Mizuno2023-px}.
The figure shows that the Ge crystal has a different trend below and above approximately 2000 channels, which corresponds to approximately 3 MeV. 
The calibration function with the combination of a linear function for the low-energy region and a quadratic function for the high-energy region was used.
GC3018 shows nonlinearity below 1.4 MeV and needs a second-order polynomial function for nonlinearity correction.
The Ge crystal of BE2820 showed negligible non-linearity below 1.4 MeV.
Figure~\ref{fig:linearity}(d) shows the residuals from the calibration using all correction functions for GX5019. 
The nonlinearity was eliminated in all energy regions below 10 MeV, and the residuals remained within one ADC channel. 

\begin{table*}[width=.75\textwidth, cols=4,pos=h]
\centering
\caption{Energy accuracy after nonlinearity correction for the three Ge detectors.
        The input dynamic range of the waveform digitizer can be selected at 0.5 and 2.0 V and the corresponding energy ranges for each detector are listed in the table.
        The result of BE2820 is only for the 0.5-V range because BE2820 is only used in low-energy regions below 2 MeV.
        The gain of the detectors is set to the maximum ($\times 10$) and the DC offset is fixed at 10\% (see Sect.~\ref{sec:Eresolution}).}
\label{tab:energy_accuracy}
\begin{tabular*}{\tblwidth}{@{}c cc cc cc@{}}
    \toprule
    \multirow{2}{*}{Input range} & \multicolumn{2}{c}{GC3018} & \multicolumn{2}{c}{GX5019} & \multicolumn{2}{c}{BE2820} \\
                & Range & Accuracy & Range & Accuracy & Range & Accuracy \\
    \midrule
    0.5 V & 2.35 MeV & 0.05 keV & 2.43 MeV & 0.06 keV & 1.55 MeV & 0.03 keV\\
    2.0 V & 9.42 MeV & 0.2  keV & 9.70 MeV & 0.3  keV & 6.22 MeV & -    \\
    \bottomrule
\end{tabular*}
\end{table*}

The selection of the dynamic range limits the energy accuracy after the correction of nonlinearity.
The input dynamic range of the digitizer can be changed to 0.5 V and 2.0 V.
The accuracy of the energy is deduced from the residuals from the calibration functions for each input dynamic range.
Table~\ref{tab:energy_accuracy} shows the accuracy of the three Ge detectors and the corresponding dynamic energy range. 
When the dynamic range becomes wider, the energy accuracy becomes worse. 

The correction of the gain drift and the pile-up effect under the high-count-rate condition is also essential for the high accuracy throughout the spectroscopy experiment when the measurement time is longer, typically more than 1 hour.
These long-term effects were phenomenologically corrected using one or two $\gamma$-ray peaks with a linear function as follows:
\begin{equation}
  \label{eq:gain_drift}
  E = A E_\mathrm{original} + N_p,
\end{equation}
where $A$ is the gain drift parameter, $E_\mathrm{original}$ is the energy before correction, and $N_p$ is the pile-up correction term.
The gain drift can be expressed by the first term 
because digital pulse processing with baseline correction does not need to consider offset. 
$N_p$ is the pile-up correction term and is 0 in low-count-rate measurements below 1~kHz, 
explained in Sect.~\ref{sec:highcountrate}. 

In summary, the correction of nonlinearity of the system and gain-drift are necessary to achieve high energy accuracy.
After the correction of nonlinearity and the long-term effect, appropriate selection of the dynamic range must be performed depending on the purpose of each measurement.
An energy accuracy smaller than 0.1 keV can be obtained in energy regions below 2 MeV and 0.3 keV for high-energy photon spectroscopy.

\subsection{Energy resolution}\label{sec:Eresolution}
Optimization of the parameters in digital pulse processing was performed to obtain the best energy resolution of the Ge detectors.
The energy resolution was deduced from the Gaussian fitting of each $\gamma$-ray peak in the energy spectrum.
A trapezoidal filter was used as an algorithm to deduce the energy in the DPP-PHA firmware.
The parameters of the trapezoidal filter are 
the flat top, pole-zero, trapezoidal rise time, peaking time, and number of samples for energy mean calculation ($N_\mathrm{sp}$)~\cite{DPP-PHAandPSD}.
In addition to these parameters, the sampling number of the baseline calculation ($N_\mathrm{sb}$), 
the input dynamic range, and the gain of the preamplifiers were examined. 
The list of the parameters and their ranges are summarized in Table~\ref{tab:Eresolution_parameters}.

$N_\mathrm{sb}$, the input dynamic range, and the gain of the preamplifiers depend on the experimental conditions.
$N_\mathrm{sb}$ should be maximized if the count rate of the measurement is lower than 2~kHz as a large $N_\mathrm{sb}$ helps reduce the effect of high-frequency noise in the pulse height calculation.
When the count rate becomes higher, $N_\mathrm{sb}$ should be changed to a lower value to prevent the pile-up effect, as explained in Sect.~\ref{sec:highcountrate}.
The input dynamic range of the digitizer and the gain of preamplifiers should be selected based on the requirement of each experiment as 
a wider energy dynamic range results in worse energy resolution.
The energy dynamic ranges with GC3018, GX5019, and BE2820 are presented in Table~\ref{tab:energy_accuracy}.

\begin{figure}
  \centering
  \includegraphics[scale=0.3]{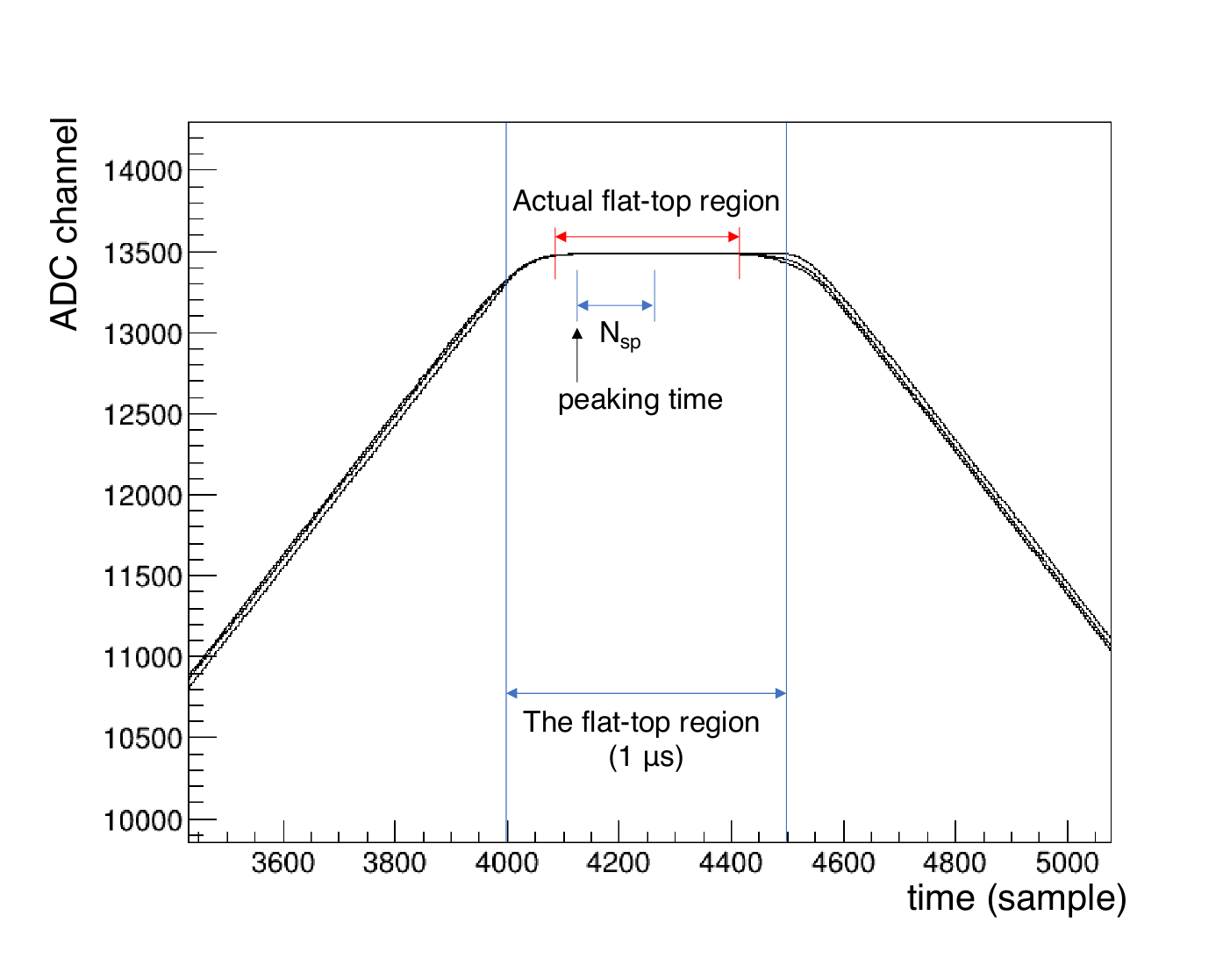}
  \caption{Trapezoidal-filtered waveforms of BE2820 zoomed in the flat-top area.
  The edges of the flat-top region show different shapes 
  with fluctuations in the shapes of the signals.}
  \label{fig:trapezoid_flattop}
\end{figure}

The parameters of the trapezoidal filter were optimized by comparing the energy resolutions.
Prior to the optimization, the pole-zero parameters 
were fixed by evaluating the output waveforms. 
The optimized parameters of the proposed system are summarized in Table~\ref{tab:Eresolution_parameters}. 
The best value of the trapezoidal rise time ranged from 5--8 $\mu$s, and the best values of the flat top ranged from 1.0--1.5 $\mu$s as usual parameters for HPGe detectors.
The trapezoidal rise time lower than 4 $\mu$s does not provide sufficient sampling numbers to reduce high-frequency noise.
The peaking time and $N_\mathrm{sp}$ correspond to the pick-up area in the flat top region of the trapezoid.
Figure~\ref{fig:trapezoid_flattop} shows the trapezoidal-filtered waveforms of BE2820. 
Each waveform had different shapes near the edges of the flat-top region, as indicated by the vertical lines.
This difference is attributed to the fluctuations in the 
shapes of the rising part of the pulses.
Taking samples of the trapezoidal height from the actual flat-top region shown in the figure is needed to obtain a better energy resolution.
Therefore, the best values of the peaking time and the $N_\mathrm{sp}$ were set to 20\% and 64 samples for BE2820, 20\% and 16 samples for GX5019, and 50\% and 16 samples for GC3018, respectively, by comparing the energy resolutions with changing these values iteratively.

\begin{table*}[width=.75\textwidth, cols=4,pos=h]
  \caption{Parameters of the proposed system. The range of the parameters is shown in the second column and the optimized parameters for obtaining the best energy resolution are listed for each detector. The parameters corresponding to the dynamic range (input dynamic range and Ge pre-amplifier gain) must be selected based on the requirement of each experiment.}
  \label{tab:Eresolution_parameters}
  \begin{tabular*}{\tblwidth}{@{}lcccc@{}}
\toprule
Parameter       &   Parameter range  & GC3018 & GX5019 & BE2820 \\
\midrule
Flat top ($\mu$s)              &  0.5 -- 2.0      & 1 & 1 & 1 \\
Pole-zero ($\mu$s)             &    40-60         & 47 & 49 & 49 \\
Trapezoidal rise time ($\mu$s) &    4-10          & 7 & 7 & 6 \\
Peaking time (\%)              &   0-100          & 50 & 20 & 20 \\
$N_\mathrm{sp}$                & 4, 16, 64        & 16 & 64 & 64 \\
$N_\mathrm{sb}$                & 1024, 4096, 16384& 16384 & 16384 & 16384 \\
Input dynamic range (V)        &   2.0, 0.5       & - & - & -\\
Ge preAmp Gain                 & $\times$1, 2, 5, 10& - & - & - \\
\bottomrule
  \end{tabular*}
\end{table*}

\begin{figure}
  \centering
  \includegraphics[scale=0.35]{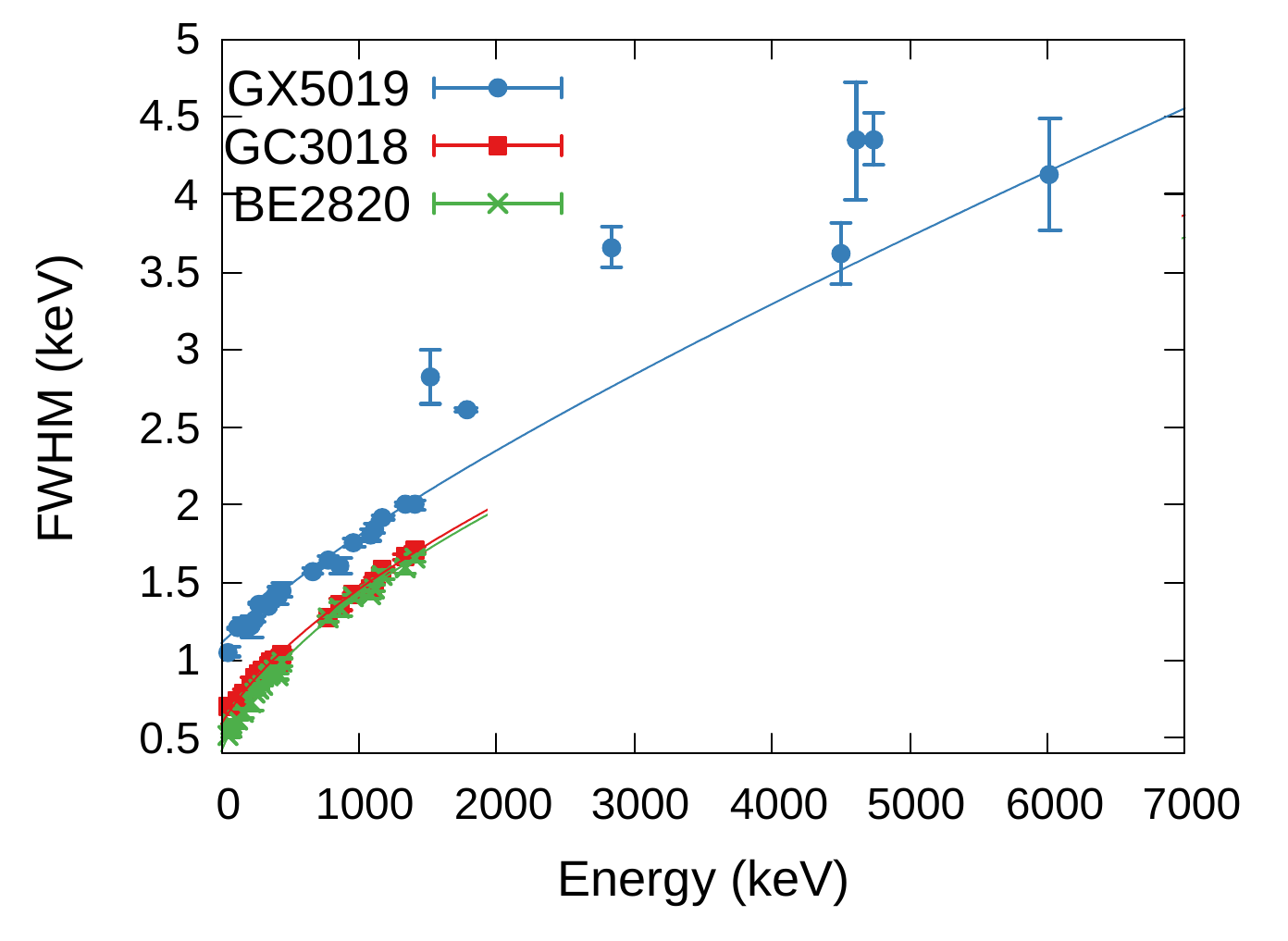}
  \caption{Energy resolutions of Ge detectors with the best parameters presented in Table~\ref{tab:Eresolution_parameters}.
  The points represent the measured values and the lines are the fitting curves obtained using Eq.~(\ref{eq:Eres_fitting}).}
  \label{fig:Eresolution}
\end{figure}

The energy resolutions of Ge detectors obtained using the best parameters are shown in Fig.~\ref{fig:Eresolution}.
The fitting functions in the figure are expressed as,
\begin{equation}
  \label{eq:Eres_fitting}
  \sigma_\mathrm{FWHM}(E) = \sqrt{aE^2+bE+c},
\end{equation}
where $a, b$, and $c$ are the fitting parameters; $E$ is the energy; and $\sigma_{\mathrm{FMHM}}$ is the energy resolution in full-width at half maximum (FWHM). 
Note that, the energy resolutions of GX5019 in high-energy regions were obtained from the in-beam measurement of the $^{27}\mathrm{Al}(p,\gamma){}^{28}\mathrm{Si}$ reaction, and they showed worse energy resolution than that of low-energy regions obtained from the offline measurement because the noise condition of beam facility was worse.

The obtained energy resolutions were the same or better than those obtained using the analog DAQ systems (using a shaping amplifier and multi-channel analyzer) and the guaranteed values in the specification sheets.
The difference in energy resolutions among the detectors originated from the crystal size and the dynamic range of each detector.

\subsection{Performance under high count-rate conditions}
\label{sec:highcountrate}

In muonic X-ray spectroscopy, several X-rays and $\gamma$-rays are emitted and the count rate of the detector is estimated to be a few kHz at PSI.
Therefore, the detector's performance, namely, a drift of the peak channel and deterioration of the energy resolution, was studied under high-count-rate conditions. 
A high count-rate condition was created 
by changing the distance ($d$) between the standard $\gamma$-ray sources and the Ge detectors from 10 cm to 42 cm.

In the DAQ system, the pile-up rejection time was set to 8 $\mu$s and signals detected in 8 $\mu$s from the previous signals were rejected.
As the time constant of the preamplifier is approximately 50 $\mu$s, 
output signals from the preamplifier require approximately 120 $\mu$s to reach the baseline. 
The pile-up effect becomes considerable under the high count-rate condition 
because the baseline calculation uses the region in which the pole-zero correction is insufficient and affected by the previous signal.
Therefore, the drift of the peak channel was observed under the high count-rate conditions.
This effect can be corrected by adding the pile-up term $N_p$ introduced in Eq.~(\ref{eq:gain_drift}).
Some calibration sources or known peaks are necessary during the measurement for conducting this correction.

\begin{figure}
  \centering
  \includegraphics[scale=0.35]{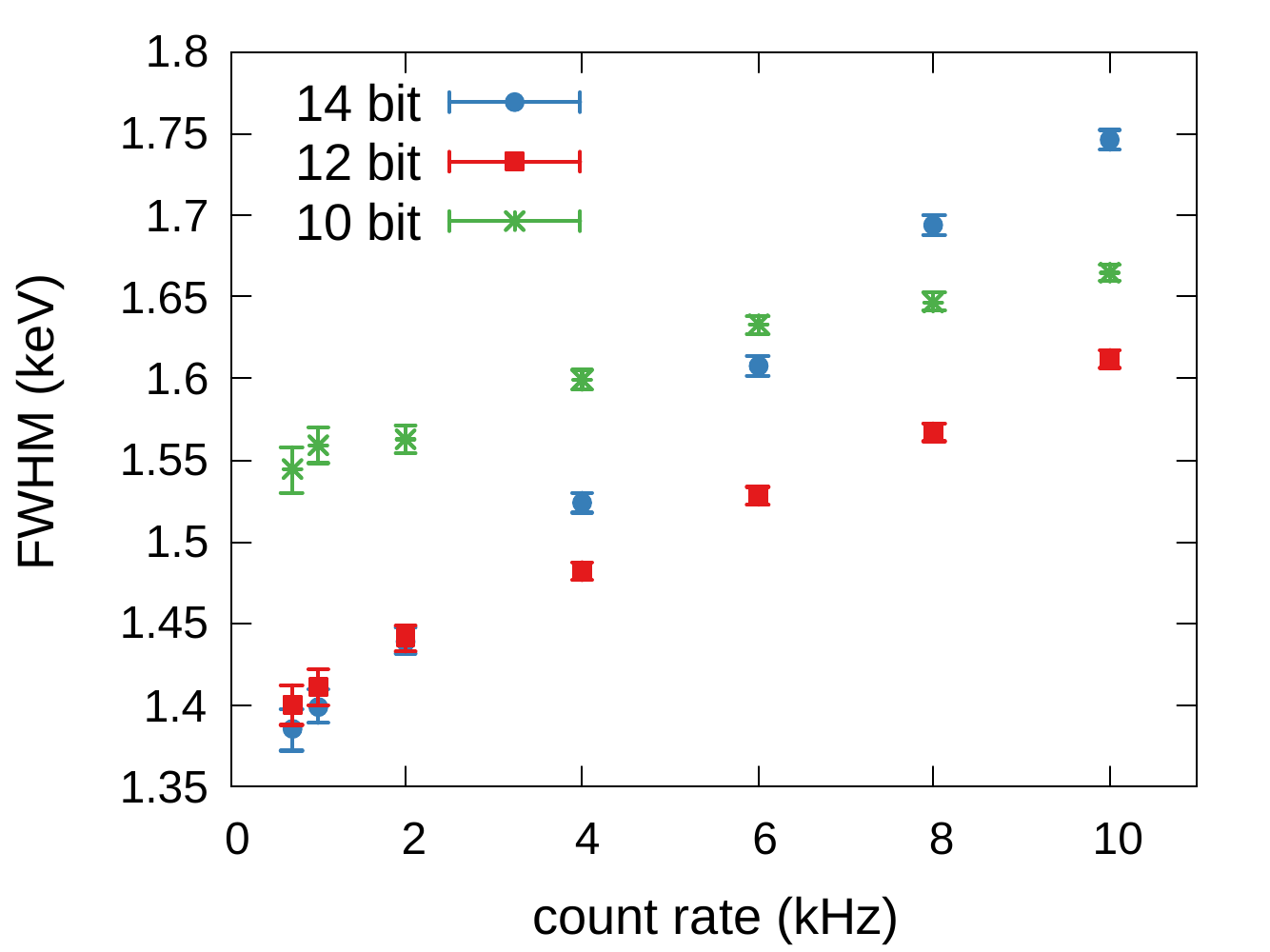}
  \caption{Energy resolution at 344 keV with GX5019 obtained by changing the count-rate condition. The sampling number of the baseline calculation ($N_\mathrm{sb}$) is set to 
  14-bit (blue circle), 12-bit (red box), and 10-bit (green cross).}
  \label{fig:Highcount_Eresolution}
\end{figure}

The pile-up effect also changes the energy resolution depending on the count rate.
The energy resolutions at 344 keV from the $^{152}$Eu $\gamma$-ray source measured using three $N_\mathrm{sb}$ values at 10-, 12-, and 14-bit for GX5019 are shown in Fig.~\ref{fig:Highcount_Eresolution}. 
The same results were obtained for GC3018 and BE2820. 
The energy resolution becomes worse as the count rate increases.
$N_\mathrm{sb}$ should be optimized based on the count rate of the measurement, namely, 14-bit below 2~kHz and 12-bit from 2--10 kHz. 
The other parameters listed in Table~\ref{tab:Eresolution_parameters} were independent of the count-rate condition.

The peaks in the energy spectra have a tail on the peak under high count-rate conditions because the baseline calculation is conducted for each signal in the digital pulse height analysis. 
Under such conditions, the response function using the Gaussian function with an exponential tail on the lower side is required. 

\subsection{Timing resolution} \label{sec:timing}
In the muonic X-ray and $\gamma$-ray spectroscopy, two subsequent events occur: muon atomic transition and nuclear muon capture~\cite{Measday2001-mw}.
These events can be distinguished by their different time scales. The muonic X rays are emitted within a few ps after muon injection, whereas the de-excitation $\gamma$-rays from the nuclear muon capture reaction are emitted in a longer time scale, typically 2 $\mu$s in the light nuclei and approximately 70 ns in the heavy nuclei.
The typical timing spectrum of the photon (X rays and $\gamma$ rays) detection is shown, for example, in Fig.~3 of Ref.~\cite{Saito2022-wl}.
The Ge detector usually has a timing resolution of approximately a few tens of ns. The timing resolution obtained using the digitizer was studied to achieve the intrinsic timing resolution of the Ge detectors.

The coincidence measurement was performed using a BaF$_2$ detector and Ge detectors using a standard $\gamma$-ray source, $^{152}$Eu.
The BaF$_2$ detector was adopted because it has a better timing resolution than the Ge detectors, and the timing resolution of the BaF$_2$ is negligible.
The timing resolution of BaF$_2$ was 4.92$\pm$0.20 ns, as deduced from a coincidence measurement using two BaF$_2$ detectors by assuming that the timing resolution of two detectors is the same.
This value is relatively worse compared with the same detector's timing resolution obtained using $^{60}$Co, 0.72$\pm$0.03 ns, 
because $\gamma$-rays from $^{152}$Sm (daughter nucleus of the $\beta^+$ decay from $^{152}$Eu) have an isomeric state with a 1.4-ns half-life. The timing resolution of 4.92$\pm$0.20 ns is sufficiently smaller enough than the Ge detector's timing resolution.
The timing resolution was evaluated for the measured $\gamma$-ray peaks by the Gaussian fitting of the timing-difference spectra between the Ge detectors and the BaF$_2$ detector.

Five timing pick-off methods were compared: RC-CR$^2$, constant fraction discriminator (CFD), amplitude and risetime compensated timing (ARC), leading edge timing (LET), and waveform analysis.
RC-CR$^2$ is a widely used timing pick-off method in digital processing because
it can obtain the timing for almost all input signals above the threshold and is easy to be implemented in FPGA~\cite{Guzik2013-zv}.
CFD, ARC, and LET are typical timing pick-off methods usually used in the analog DAQ systems of the Ge detectors.
CFD and ARC can obtain the timing independently with the pulse height, whereas LET shows the energy dependence, called ``time walk''. 
The difference between the CFD and the ARC is the delay time in their algorisms; delay times of 150 ns and 50 ns for the CFD and ARC, respectively, were used in this study. 
RC-CR$^2$ was implemented in the DPP-PHA firmware, and the CFD, ARC, and LET were obtained using the DPP-PSD firmware. 
The waveform obtained using the flash ADC was analyzed offline and the pulse arrival timing was evaluated, which is hereafter referred to as the waveform analysis method.
In the offline waveform analysis method, 
the time when the pulse height becomes a small fraction (GC3018: 1/100, GX5019: 1/300, BE2820: 1/500) of the full pulse height is picked off.
The waveform analysis result is used as the reference value over the results obtained using other FPGA-based methods because the waveform analysis can obtain a timing resolution close to the intrinsic timing resolution of the Ge detectors. 
The waveform analysis is only applicable in low-count-rate measurements with the present system because this method requires large data transfers.

\begin{figure}
  \centering
  \includegraphics[scale=0.26]{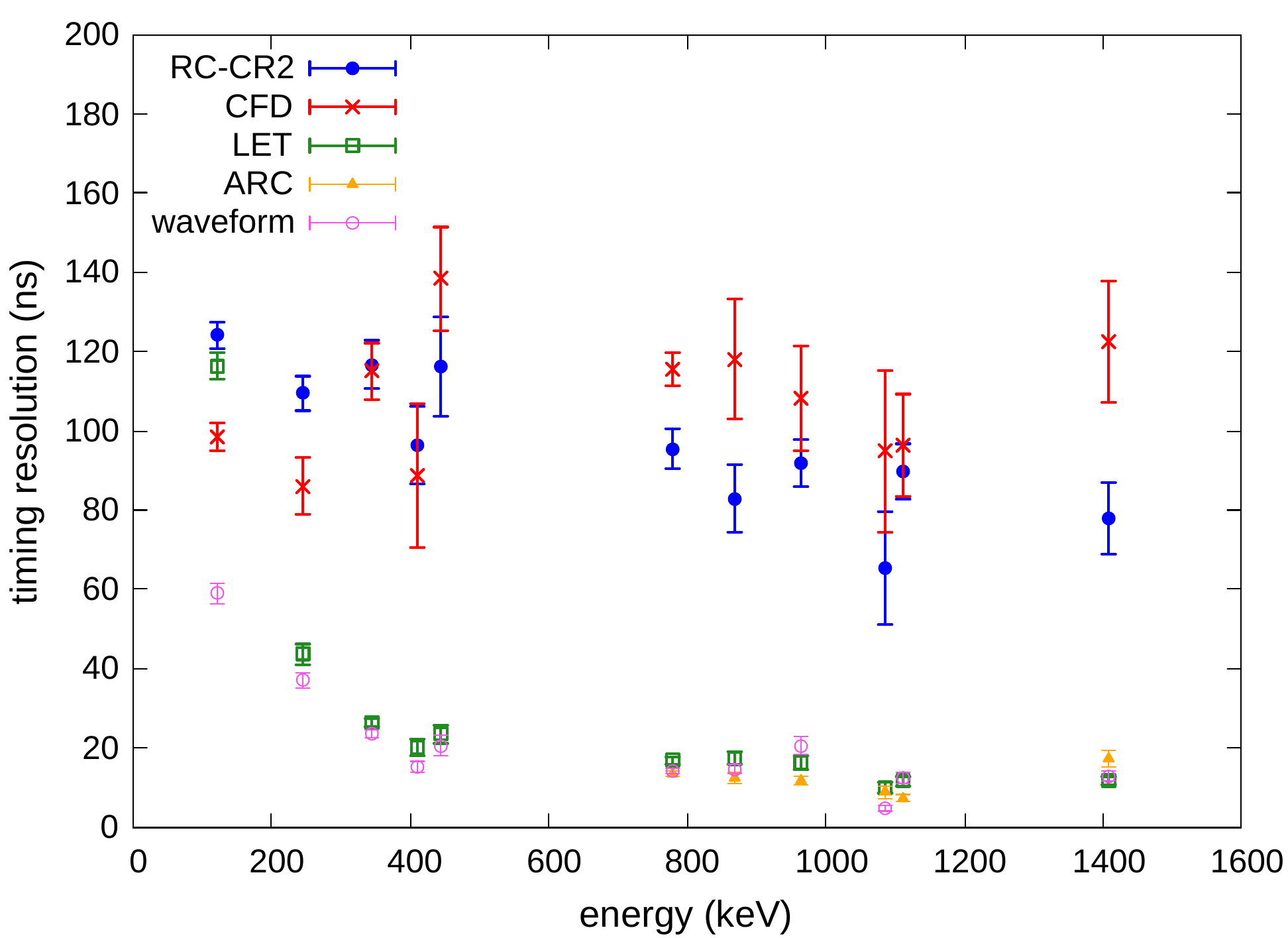}
  \caption{Timing resolution of GX5019 obtained using CR-RC$^2$, CFD, ARC, LET, and waveform analysis.
  The fraction and delay time of CFD and ARC are 50\%-150 ns and 50\%-50 ns, respectively.}
  \label{fig:Tresolution}
\end{figure}

The timing resolutions of GX5019 obtained using the five timing pick-off methods are shown in Fig.~\ref{fig:Tresolution}.
RC-CR$^2$ and CFD showed worse timing resolutions in wide energy regions, and ARC and LET showed a timing resolution of approximately 10--20 ns in energy regions above 400 keV.
LET showed a timing resolution almost similar to that obtained using waveform analysis; however, the timing resolution of LET was worse in energy region below 100 keV.

\begin{figure}
  \centering
  \includegraphics[scale=0.28]{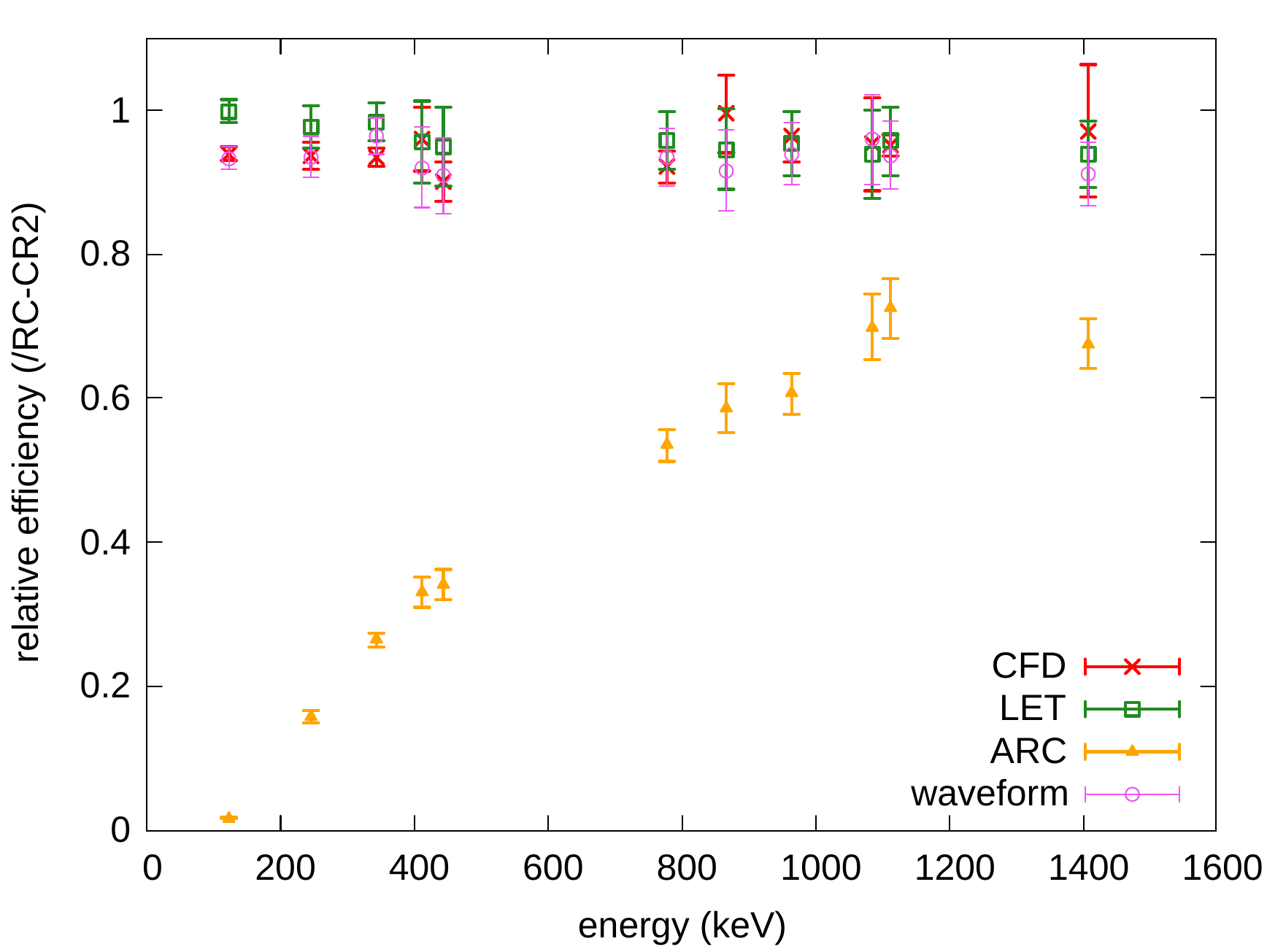}
  \caption{Number of processed pulses obtained using the following timing pick-off methods: CFD, ARC, LET, and waveform analysis, which were normalized with those obtained using RC-CR$^2$.}
  \label{fig:ARC_efficiency}
\end{figure}

The ratios of the lost pulses in each timing pick-off method were also compared. 
The numbers of appropriately processed pulses obtained using the CFD, ARC, LET, and waveform analysis are shown in Fig.~\ref{fig:ARC_efficiency}, which are normalized with those obtained using RC-CR$^2$. 
CFD, LET, and waveform analysis results were almost the same as those obtained using RC-CR$^{2}$, whereas ARC lost some parts of the signals, especially in the low-energy region.
Thus, the timing resolution of ARC is not shown in Fig.~\ref{fig:Tresolution} for low-energy regions. 
ARC needs sufficient pulse height before the delay time to be distinguished from noise. 

\begin{figure}
  \centering
  \includegraphics[scale=0.4]{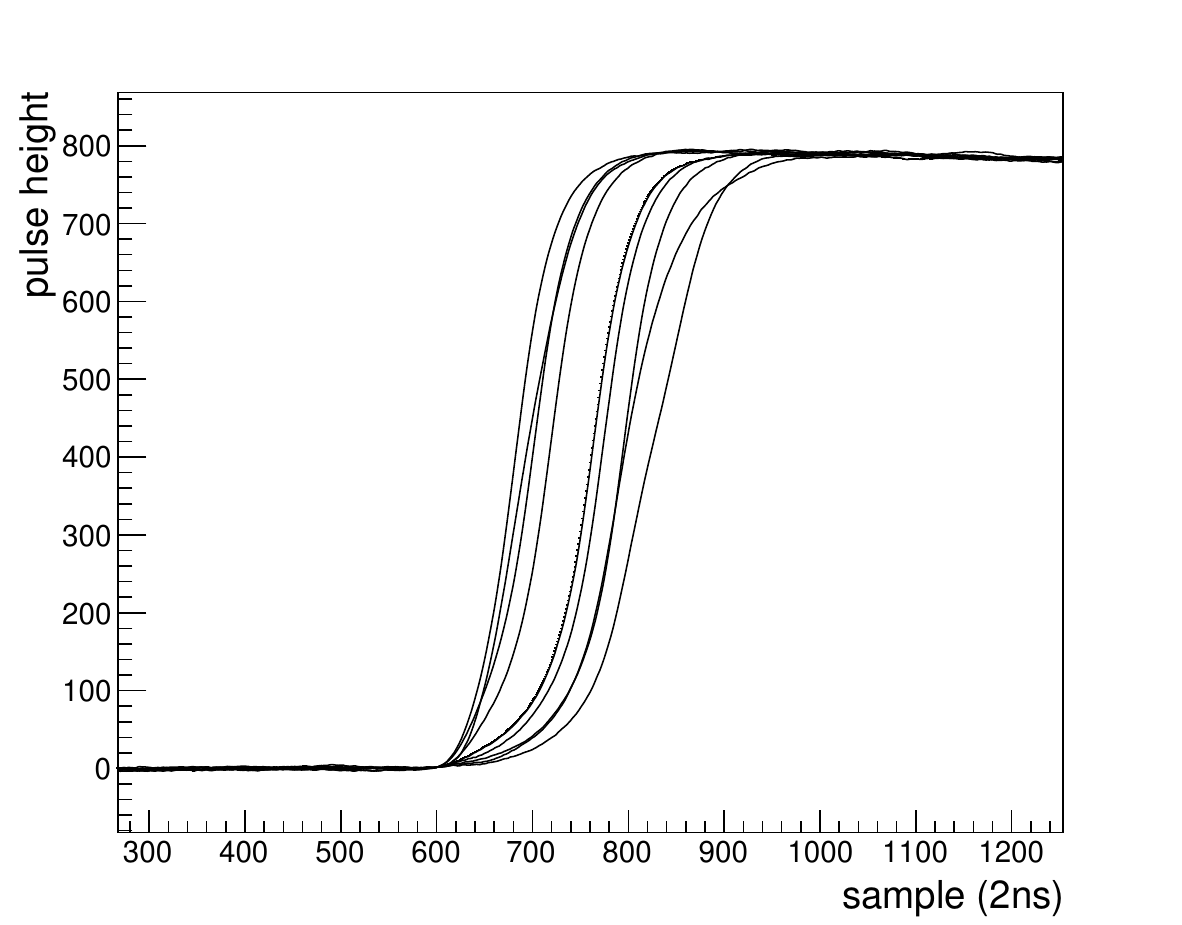}
  \caption{Typical waveforms with the same pulse height and 
  aligned with the time calculated using the waveform analysis of BE2820.}
  \label{fig:rise_shape}
\end{figure}

The differences in timing resolutions and the numbers of processed pulses among the timing pick-off methods primarily originated from 
the differences in the rise time and shape of the waveform, as shown in Fig.~\ref{fig:rise_shape}, which are the typical waveforms of BE2820.
The difference in the interaction position of the photon inside the Ge crystal leads to a difference in the rise time and shape of the waveform. 
The performance of the preamplifier characterizes the effect of position dispersive on the difference in the rising shape. 
As RC-CR$^2$ and CFD obtain the timing of the constant ratio of the pulse height, their performances are strongly affected by the difference in rise times and rising waveforms, which result in a worse timing resolution.

Based on these results, we showed that LET can be used to determine the best timing resolution, which is almost similar to the timing resolution obtained using the waveform analysis in energy regions of approximately more than 200 keV. 
When a higher timing resolution is required in the low-energy region below 200 keV, the offline waveform analysis should be used for the timing pick-off method. 
Note that, the correction of the time walk is necessary when timing information obtained with LET is compared with that of other detectors, such as coincidence analyses.

\subsection{The photopeak efficiency}
\label{sec:efficiency}

\begin{figure}
  \centering
  \includegraphics[scale=0.35]{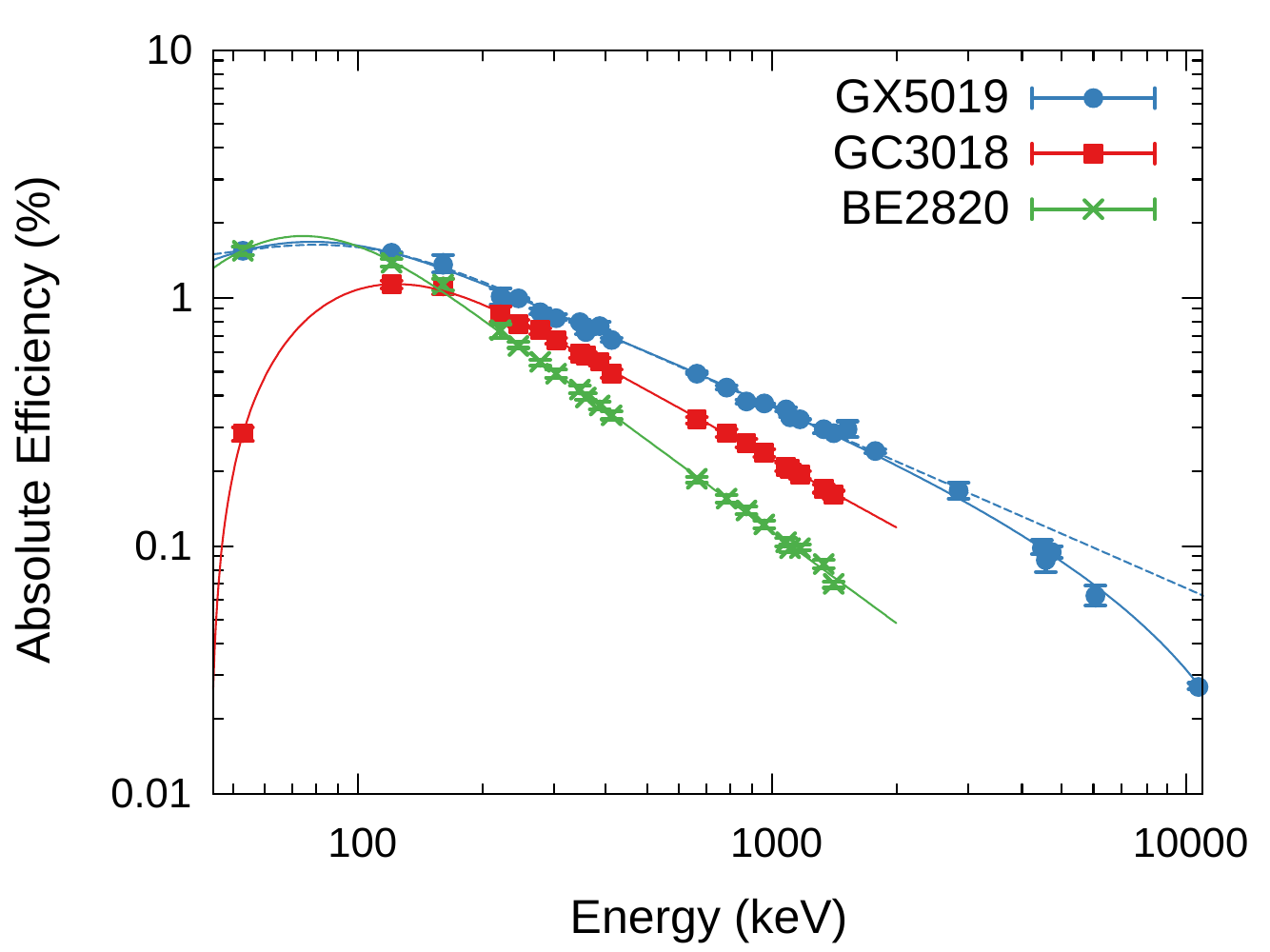}
  \caption{Photopeak efficiencies of GC3018, GX5019 and BE2820. 
  The results of GC3018 and BE2820 are measured using $^{60}$Co, $^{133}$Ba, and $^{152}$Eu.
  The result of GX5019 is the result of high-energy $\gamma$-ray measurement with the p+Al reaction and smoothed in the low-energy region. The distance of the Ge detectors from the target was 10 cm. The dotted line shows the extrapolation line from the energy region below 1.5 MeV for GX5019, obtained using Eq.~(\ref{eq:lowE_eff}).}
  \label{fig:Efficiency_measure}
\end{figure}

The photopeak efficiency of the Ge detectors was measured with standard $\gamma$-ray sources and in-beam $\gamma$-ray measurements using the ${}^{27}\mathrm{Al}(p,\gamma){}^{28}\mathrm{Si}$ reaction.
Figure~\ref{fig:Efficiency_measure} shows the photopeak efficiencies of each Ge detector placed 10 cm apart from the sources/target.
The efficiencies of GC3018 and BE2820 were measured using $^{60}$Co, $^{133}$Ba, $^{137}$Cs, and $^{152}$Eu.
The efficiency of GX5019 is the result of standard $\gamma$-ray measurements and in-beam $\gamma$-ray measurements. 
Two functions,
\begin{align}
  \label{eq:lowE_eff}
  \epsilon =& a_0E^{-a_1}-a_2\mathrm{exp}(-a_3E) \\
  \label{eq:highE_eff}
  \epsilon =& a_0E^{-a_1}-a_2\mathrm{exp}(-a_3E)-a_4\frac{(E-a_5)}{\sqrt{a_6+(E-a_5)^2}}
\end{align}
were used as fitting functions in the figure. Eq.~(\ref{eq:lowE_eff}) represents the efficiency curve for GC3018 and BE2820 and
Eq.~(\ref{eq:highE_eff}) represents the efficiency curve for GX5019, where $\epsilon$ is the efficiency and $a_0$--$a_6$ are the fitting parameters. 
As shown in Fig.~\ref{fig:Efficiency_measure}, the efficiency curve of GX5019 shows a sharp turn at approximately 3 MeV.
Therefore, the measured efficiency value differed from the value deduced from extrapolation using Eq.~(\ref{eq:lowE_eff}) in energy regions below 1.5 MeV, as shown in the dotted line in Fig.~\ref{fig:Efficiency_measure}. 
Similar turns at approximately 2--3 MeV of the photo-peak efficiency are previously reported~\cite{Singh1971-as,McCallum1975-fy,Molnar2002-le,Elekes2003-db} and explained using a model, including pair production and multiple processes~\cite{Hajnal1974-ry}.

\begin{figure}
  \centering
  \includegraphics[scale=0.35]{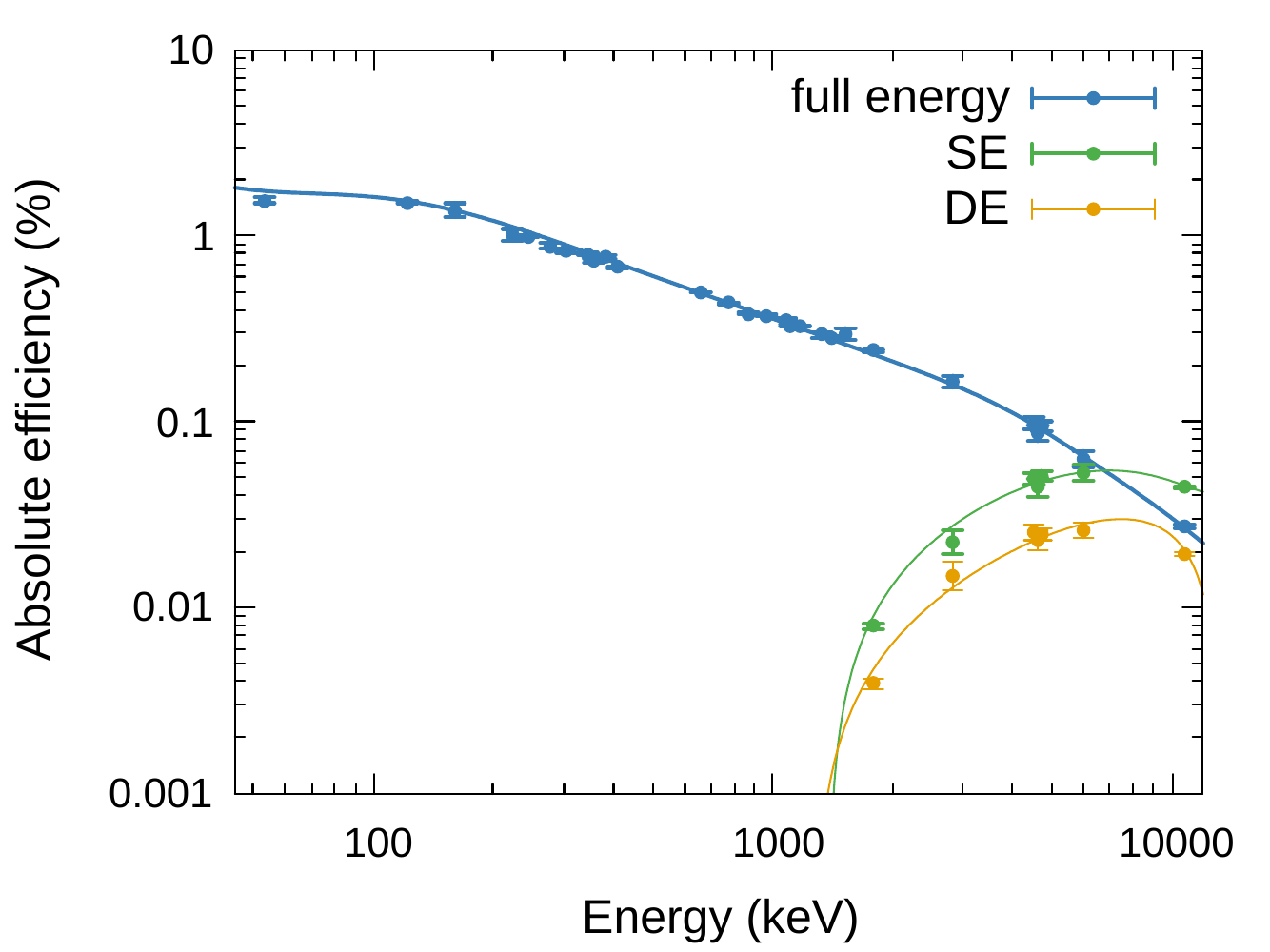}
  \caption{Efficiency of GX5019 obtained using the measurement and corrected GEANT4 simulation.
  The green and yellow points represent the efficiencies of the SE and DE peaks (circle: measurement; line: GEANT4).}
  \label{fig:Efficiency_g4}
\end{figure}

The GEANT4 simulation was performed for the reproduction of the efficiency curve.
Figure~\ref{fig:Efficiency_g4} shows the result of measurements and the GEANT4 simulation of GX5019 for the full energy, single escape (SE), and double escape (DE) peaks. 
The GEANT4 simulation overestimated the efficiencies because of the inaccuracy of the insensitive volume in the Ge crystal and the uncertainty of the charge collection~\cite{Hurtado2004-lz,CebastienJoel2018-tr}. 
This overestimation is independent of the energy and the distance between the target and the detector and can be corrected with a constant factor.
The measured efficiencies were reproduced by the results obtained using GEANT4 divided by the correction factors: 1.172(35) for GC3018, 1.071(32) for GX5019, and 1.086(33) for BE2820.
The corrected efficiency with constant values obtained using GEANT4 reproduced the measured values well, including the SE and DE peaks, as shown in Fig.~\ref{fig:Efficiency_g4}.
The efficiency kink observed in this study can be understood based on the escape of the energy accompanied by the pair creation and multiple processes.

The measured efficiency curve suggests that the extrapolation from the measured values using standard sources below 2 MeV is insufficient. 
Therefore, we proposed the use of muonic X-rays from Au and Bi as efficiency standards in high-energy regions at the muon facility, as explained in Sect.~\ref{sec:MIXE}.
\subsection{Compton suppressor}\label{sec:compton}

\begin{figure}
  \centering
  \includegraphics[scale=0.43]{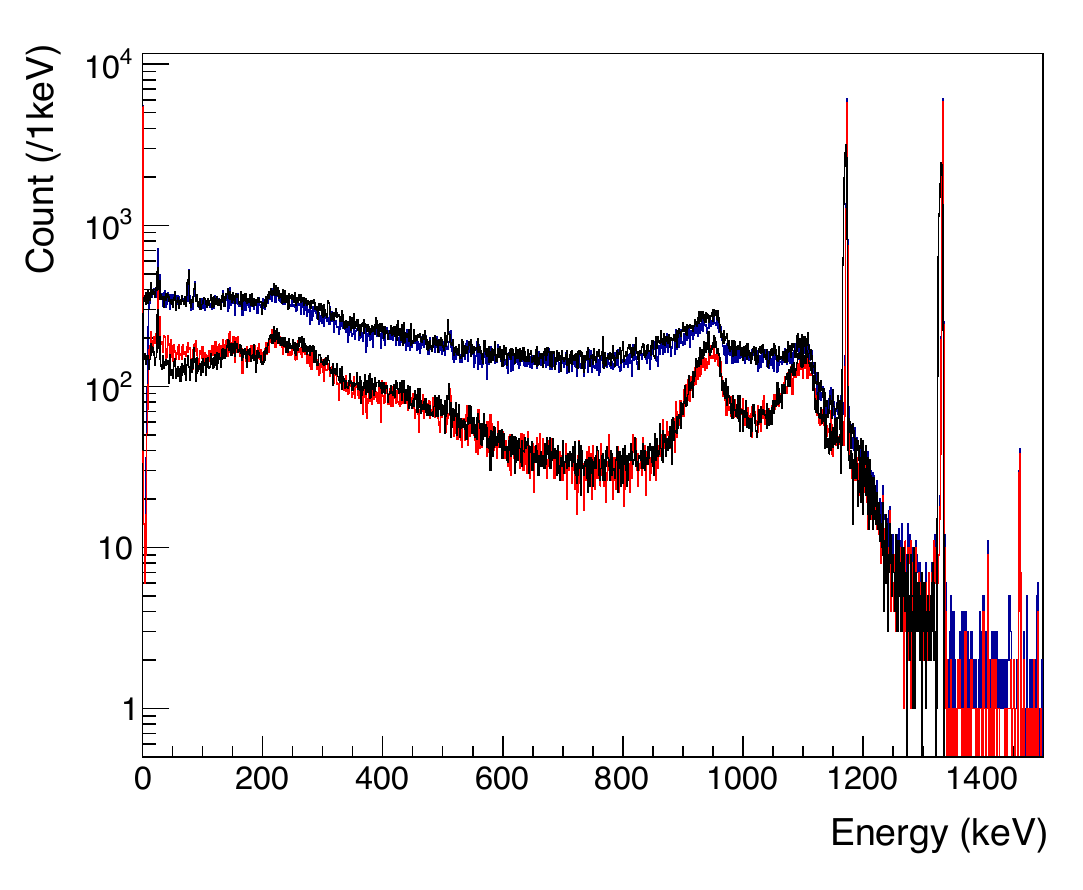}
  \caption{Energy spectra with (the blue line) and without (the red line) Compton suppression analysis of $^{60}$Co. Black lines represent the spectra calculated using GEANT4.}
  \label{fig:Compton_suppressed_lowE}
\end{figure}

\begin{figure}
  \centering
  \includegraphics[scale=0.43]{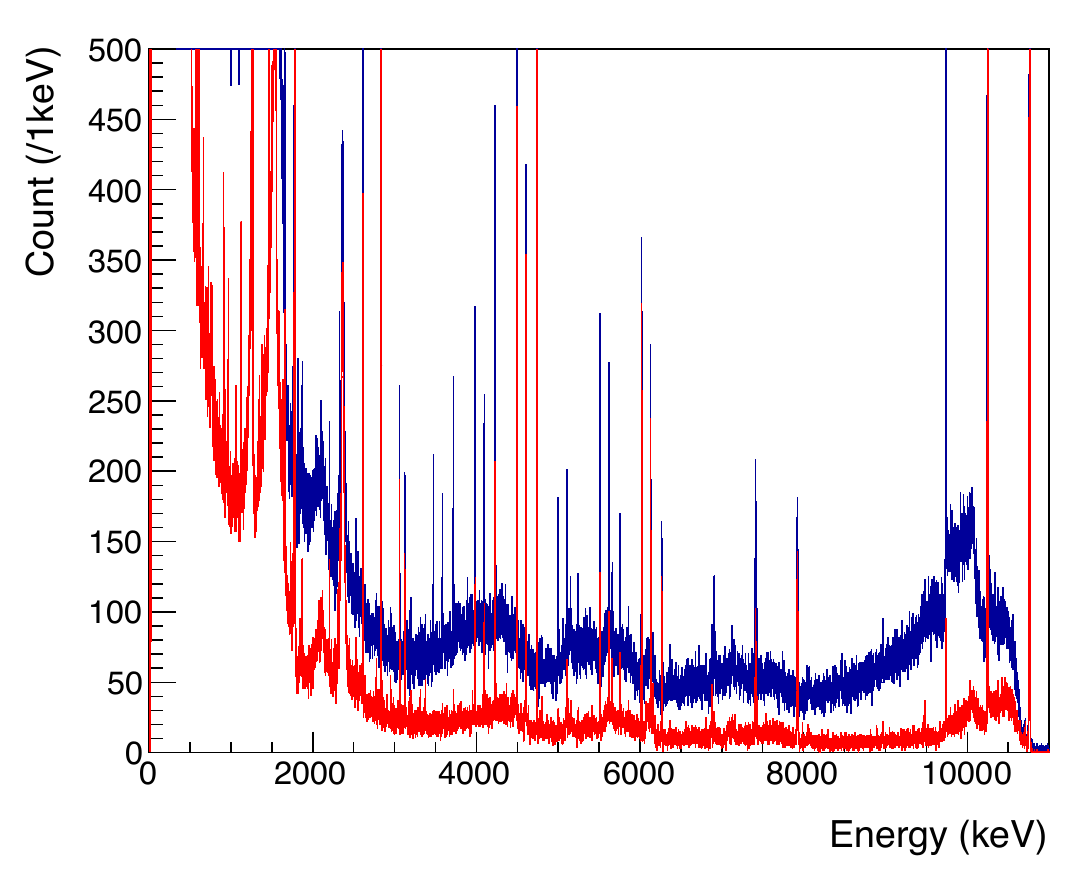}
  \caption{Energy spectra with (the blue line) and without (the red line) Compton suppression analysis of the high energy $\gamma$-ray measurements using $^{27}\mathrm{Al}(p,\gamma){}^{28}\mathrm{Si}$ reaction. 
  The top of the peaks are omitted.}
  \label{fig:Compton_suppressed_highE}
\end{figure}

Figures~\ref{fig:Compton_suppressed_lowE} and \ref{fig:Compton_suppressed_highE} show the energy spectra of the Ge detector with and without BGO Compton suppression analysis obtained using BE2820.
The blue (above) line represents the original spectrum, and the red (below) line represents the spectrum obtained by taking anti-coincidence with BGO detectors.
The Compton component is suppressed by approximately 50--60\% below the Compton edge 
by maintaining the full energy peak counts of more than 95\% in all the energy range below 11 MeV. 
The peak count of SE and DE were also reduced as shown in Fig.~\ref{fig:Compton_suppressed_highE}. 
The spectrum around the Compton edge is not reduced by the Compton suppressors 
because the BGO crystals are placed to suppress the forward scattering to reduce the Compton component in the relatively low-energy regions.



The performance of the Compton suppressor is also simulated in GEANT4.
The black line in Fig.~\ref{fig:Compton_suppressed_lowE} shows the result of the GEANT4 simulation 
normalized by the peak count with the measurements, which reproduced the measured spectrum 
of the standard source well.
Thus, GEANT4 simulation can be used for designing the geometries of the detection system 
to determine the performance of Compton suppression around the objective peak of each measurement.

\section{Muonic X-ray spectroscopy} \label{sec:MIXE}

The performance of the detector system in the MIXE experiment was demonstrated at PSI.

The muonic X-rays of a typical carbonaceous meteorite, an Allende meteorite, were measured to evaluate the performance of BGO Compton suppressors in the MIXE experiments.
The Allende meteorite contains a low composition of carbon (0.3~wt\%) and a high composition of magnesium (15~wt\%), silicon (16~wt\%), and iron (24~wt\%)~\cite{Mason1975-ko, Stracke2012-vw}.
The energy of muonic K$_\alpha$ X-rays of carbon is 75 keV and those of magnesium, silicon, and iron are 297, 400, and 1255 keV, respectively. 
Therefore, the energy peaks of K$_\alpha$ X-rays emitted from carbon in the spectrum are on the top of the Compton component of heavier elements and are difficult to identify when the carbon composition is small~\cite{Terada2014-cw}.
However, the use of Compton suppressors improves the S/N ratio under such conditions.
Hence, the S/N ratio of muonic X-rays emitted from carbon 
of the Allende meteorite was measured with and without Compton suppression.

Furthermore, the muonic X-ray spectroscopy for $^{197}$Au, $^{208}$Pb, and $^{209}$Bi was conducted to obtain the calibration reference data for high-energy photons. 
The energy and efficiency of the prototype detector of GX5019 were evaluated with standard gamma-ray sources and an in-beam measurement using the $^{27}\mathrm{Al}(p,\gamma){}^{28}\mathrm{Si}$ reaction, as explained in Sect.~\ref{sec:Performance}.
As the in-beam measurement for calibrating high-energy photons cannot be conducted at the muon beam facility, general references 
for high-energy photon calibration are required for performing in-situ calibrations.
The muonic X-rays of the high $Z$ element can be used as the reference, as they have a high energy of approximately 6 MeV. 
$^{197}$Au and $^{209}$Bi are the best candidates for the reference target because they have the highest $Z$ among the stable nuclei and are enriched in natural composition. $^{208}$Pb is also a candidate for the reference target; however, the preparation of an enriched and flat-thickness target is expensive.

The measurements of muonic X-ray energies and intensities of $^{197}$Au, $^{208}$Pb, and $^{209}$Bi have been conducted in several studies.
The muonic X-ray energies of $^{208}$Pb have been intensively measured with high precision in the context of improving the model for deducing the nuclear charge radius~\cite{Anderson1966-ir, Anderson1969-fr, Powers1968-rq, Backenstoss1970-wj, Kessler1975-sw, Bergem1988-nf, Hoehn1984-op}, and the energy peaks of muonic X rays from $^{208}$Pb are often used as the energy reference at muon facilities. 
The relative intensities of each X-ray have also been measured~\cite{Anderson1969-fr}, which are used for efficiency calibration.
For $^{197}$Au, measurements of the muonic X-ray energies~\cite{Acker1966-kh, Powers1974-aa, Measday2007-zh} and intensities~\cite{Hartmann1982-wi, Measday2007-zh} have been reported.
The measurement of X-ray energies from $^{209}$Bi are reported in Refs.~\cite{Acker1966-kh, Powers1968-rq, Bardin1967-py, Engfer1974-km, Schneuwly1972-cv, Measday2007-zh} and corresponding intensities are reported in \cite{Measday2007-zh}.
However, in these measurements, the efficiencies of the photon detectors were deduced by the extrapolation from the standard $\gamma$-ray sources with energies below approximately 1.5 MeV, and this method may overestimate the efficiencies in high-energy regions, as explained in Sect.~\ref{sec:efficiency}.
In addition, the relative intensities of muonic X-rays from $^{197}$Au and $^{209}$Bi were reported with an assumption on the probability of nonradiative transitions.
Therefore, the energies and relative intensities of muonic X-rays (K, L, M, and N series) from $^{197}$Au, $^{208}$Pb, and $^{209}$Bi were measured to obtain accurate and useful data for photon detector calibration at muon facilities.

The experiment was performed at the $\pi$E1 beamline of the high-intensity proton accelerator (HIPA) at PSI~\cite{Kiselev2021-se} in Switzerland.
The typical muon rate was approximately 20~kHz at a momentum of 30 MeV/c for the meteorite target, and approximately 39~kHz at a momentum of 40 MeV/c for the metal targets.
The momentum bite of the beam ($\Delta p/p$) was 2\%. 
The beam spot was collimated to 18 mm in diameter.
Two Ge detectors, GX5019 and BE2820 with Compton suppressors, were set at an angle of 60 and $-$120 degrees with respect to the beamline and at a distance of 10 cm and 20.3 cm from the targets, respectively.
The DAQ system was the same as that explained in Sect.~3.

The Allende meteorite sample was shaped such that resembled approximately 4-cm square with 5-mm thickness and irradiated with a muon beam for 2.5 h. The sample was held using an aluminum target holder to avoid the background of carbon from the target holder. Note that, the contaminations of nitrogen and oxygen from the air were not avoided with this setup.

The $^{197}$Au and $^{209}$Bi metal targets were square-shaped and 50 mm in length on each side with a thickness of 0.482 mm and 1.07 mm, respectively.
$^{197}$Au target with the thickness of 0.742~mm 
(additional $\phi$48 mm, 0.26-mm thick disk was placed on the square $^{197}$Au target) was also used to evaluate the effect of self-absorption of the X-rays.
The measurement times were 1~h for 0.482-mm and 0.742-mm thick $^{197}$Au each and 1.5~h for 1.07-mm thick $^{209}$Bi.
The $^{208}$Pb was approximately 0.8-mm thick and used to evaluate the energy calibration method, as explained in Sect.~\ref{sec:linearity}.
The standard $^{60}$Co source was placed around the target during the measurement to evaluate the gain drift.

The energy and efficiency of the detector system were calibrated, as discussed in Sect.~3.
The effect of the gain drift and pile-up during the measurements was corrected using Eq.~(\ref{eq:gain_drift}). 
In addition to a 511-keV peak and $\gamma$-ray peaks from $^{60}$Co, the peaks of 1173 keV and 1332 keV, 266 keV (deexcitation of $^{205}$Tl created by the nuclear muon capture reaction of $^{208}$Pb), and 1809 keV (background $\gamma$ rays from $^{27}$Al) were used as references for the long-term correction of $^{208}$Pb measurements. 
For the long-term correction of the $^{209}$Bi target, peaks of 277 and 2614 keV from $^{208}$Pb created by the muon capture reaction of $^{209}$Bi were used as additional reference peaks. 
Only four peaks at 511, 1173, 1332, and 356 keV(deexcitation of $^{196}$Pt) were used for $^{197}$Au.
For $^{197}$Au, the uncertainty of the pile-up effect correction became large in the energy region above 1.3 MeV because of a lack of known energy peaks in high-energy regions. 

The photo peak efficiency was evaluated using the GEANT4 simulation.  
The effects of self-absorption of the X-rays in the targets were included in the simulation.
The stopping position of the muon was simulated under the same condition as the measurement, and the peak efficiencies of X-rays at each stopping position were calculated.
The peak efficiency curves of the Ge detector were evaluated for each target. 
The accuracy of self-absorption correction was evaluated by comparing the intensity results of 0.48-mm and 0.74-mm thick $^{197}$Au targets. 
\subsection{Results and Discussion: Allende meteorite}
\begin{figure}
    \centering
    \includegraphics[scale=0.6]{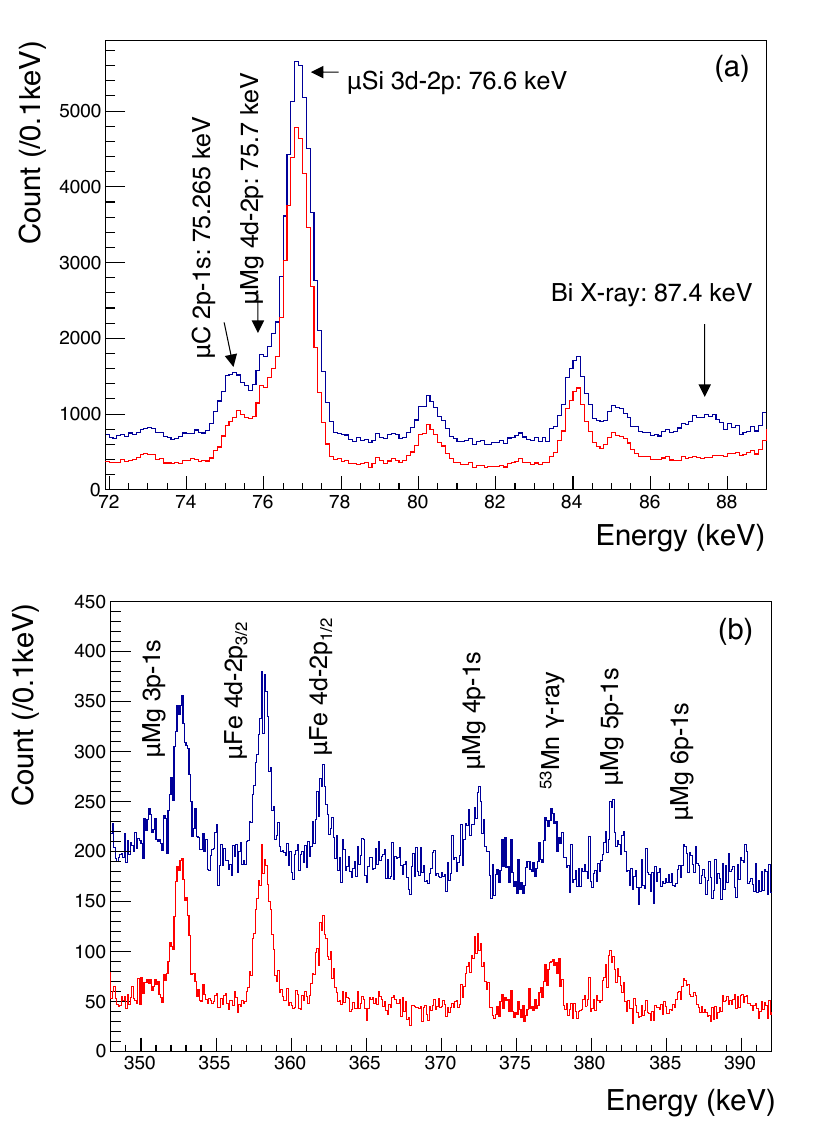}
    \caption{Energy spectra of muonic X rays obtained using BE2820 from the Allende meteorite in the energy regions of muonic K$_\alpha$ rays (2p-1s) from carbon (a) and muonic L$_\beta$ rays (4d-2p) from iron (b). The blue (above) line represents the original spectrum of the Ge detector, and the red (below) line represents the subtracted spectrum with BGO Compton suppression. In addition to the muonic L$_\beta$ rays from iron (358.2 and 362.0 keV), a part of the muonic K series from magnesium (352.6, 372.3, 381.4, and 386.3 keV) and $\gamma$-ray from $^{53}$Mn (377.8 keV) are also observed in (b).}
    \label{fig:Allende_Carbon}
\end{figure}

Figure~\ref{fig:Allende_Carbon} shows the spectrum of muonic X-rays obtained using BE2820 for the Allende meteorite in the energy regions of the muonic $K_\alpha$ ray of carbon (a) and muonic $L_\beta$ rays of iron (b). 
The blue and red lines indicate the spectrum without and with Compton suppressors, respectively. The peaks in Fig.~\ref{fig:Allende_Carbon}(a) in the 75-78 keV region are the $K_\alpha$ rays (75.265 keV) from carbon, $L_\beta$ rays (75.7 keV) from magnesium, and $L_\alpha$ rays (76.6 keV) from silicon. 
As shown in Fig.~\ref{fig:Allende_Carbon}, the Compton component was reduced by maintaining the peak component at more than 95\%.
The energy peaks of electric X-rays from bismuth contained in the BGO crystals were eliminated through subtraction using their signals, as shown in the peak in the 87.4-keV region in Fig~\ref{fig:Allende_Carbon}(a). 
The S/N ratio of the $K_\alpha$ rays originating from carbon was improved from 0.532(34) to 0.995(54) using the BGO Compton suppressor. 
Similarly, the S/N ratios of the $L_\beta$ rays from iron were improved from 0.75(7) and 0.34(6) to 2.8(4) and 1.4(2), respectively, which are approximately 3.9 times the original values. 

Based on these results, we showed that the photon detection system using the thin planer Ge detector with a Compton suppressor can detect at least 0.3~wt\% carbon 
mixed with 15~wt\% magnesium, 16~wt\% silicon, and 24~wt\% iron, which could not be detected in previous study~\cite{Terada2014-cw}.

\subsection{Results and Discussion: muonic X-rays from $^{197}$Au, $^{208}$Pb, and $^{209}$Bi}
\begin{figure*}[width=0.85\textwidth, cols=4,pos=h]
    \centering
    \includegraphics[scale=0.8]{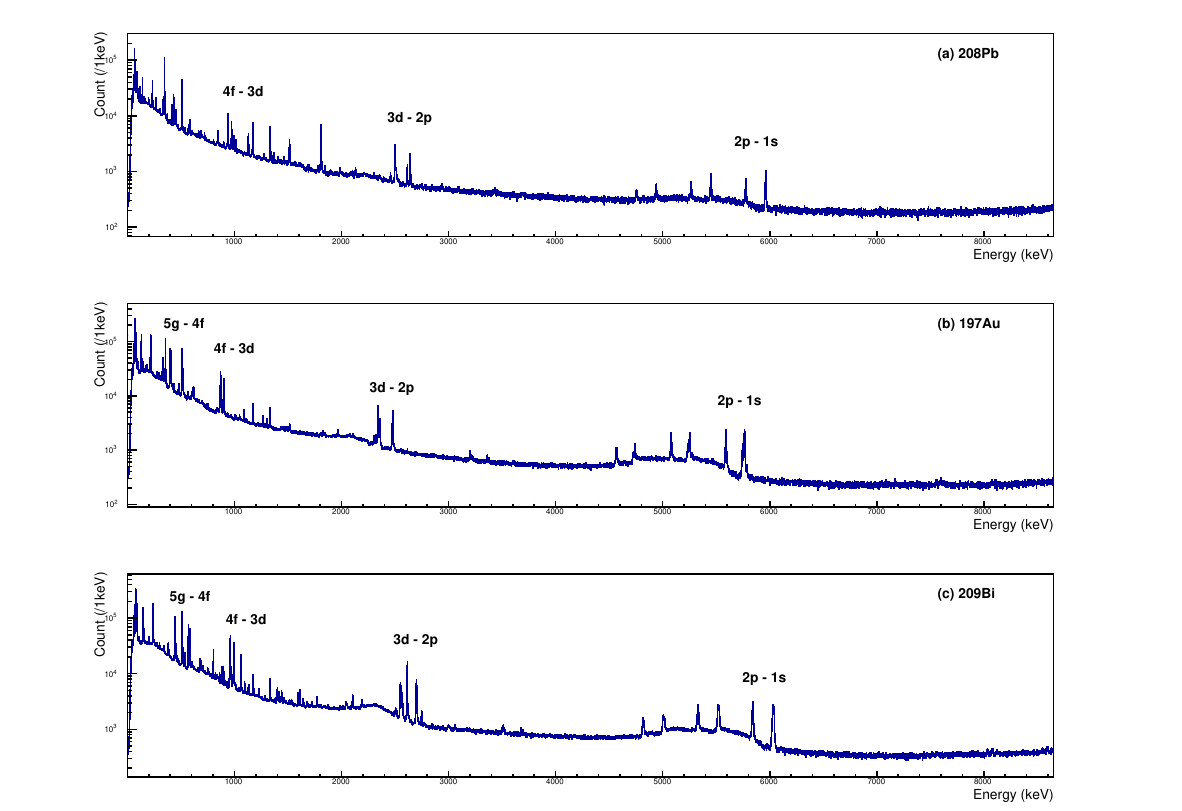}
    \caption{Energy spectrum of the muonic X-rays of $^{208}$Pb (a), $^{197}$Au (b), and $^{208}$Bi (c). X-rays from the K, L, M, and N series are identified. The SE and DE peaks are shown in the spectra.}
    \label{fig:muX_energyspectrum}
\end{figure*}

The energy spectra of muonic X-rays from $^{197}$Au, $^{208}$Pb, and $^{209}$Bi are shown in Fig.~\ref{fig:muX_energyspectrum}. 
Each peak in the energy spectra was fitted with the response function.
The measured energies and intensities of muonic X-rays of $^{208}$Pb, $^{197}$Au, and $^{209}$Bi are presented in Table~\ref{tab:muX_Pb}, \ref{tab:muX_Au}, and \ref{tab:muX_Bi}, respectively.
The literature values from the previous measurement are listed in these tables. 
Assignments of transitions are based on the energies compared with the calculated values using MuDirac~\cite{Sturniolo2021-kw}. 
$^{197}$Au and $^{209}$Bi showed hyperfine splitting owing to their nonzero nuclear spin.
The energies are listed in the tables for each hyperfine transition if they are clearly separated with the energy resolution of the Ge detector (approximately 7.4 keV in FWHM at 6.0 MeV evaluated from $^{208}$Pb measurements).
If the peaks are overlapped, the listed energies are the mean values of each peak containing several levels of hyperfine splitting.
The relative intensity of $^{208}$Pb is not shown because the thickness of the $^{208}$Pb target was not uniform, and the estimation of self-absorption using the GEANT4 simulation was insufficient for the efficiency calibration. 
The results of relative intensities of 0.48-mm and 0.74-mm thick $^{197}$Au targets were consistent in all the energy ranges, and the accuracy of self-absorption correction was confirmed. The intensities shown in Table~\ref{tab:muX_Au} are the result of 0.48-mm thick $^{197}$Au target. 
The relative intensities shown in Table~\ref{tab:muX_Au} and \ref{tab:muX_Bi} were normalized such that the largest intensity of $4f_{7/2}-3d_{5/2}$ transitions\footnote{The $4f_{7/2}-3d_{5/2}$ transitions contain several unresolved hyperfine splittings.} is 100.
The previous research reported intensities of 47.9 (12)\% for $^{197}$Au from $4f_{7/2}-3d_{5/2}$ transition~\cite{Measday2007-zh} and 75.6 (15)\% from the sum of $4f_{5/2}-3d_{3/2}$ and $4f_{7/2}-3d_{5/2}$ transitions~\cite{Hartmann1982-wi}.
For $^{209}$Bi, an intensity of 42.7 (29)\% from $4f_{7/2}-3d_{5/2}$ transition~\cite{Measday2007-zh} and 
72 (3.6)\% from the sum of $4f_{5/2}-3d_{3/2}$ and $4f_{7/2}-3d_{5/2}$ transitions~\cite{Backenstoss1970-wj} were reported, and the latter values are more reliable according to Ref.~\cite{Measday2007-zh}. 
These values are based on assumptions or model calculations, as discussed below. 
To convert to absolute intensities from the values presented in Table.~\ref{tab:muX_Au} and \ref{tab:muX_Bi}, one must multiply the values with 0.44(4) and 0.42(4), respectively.

The muonic X-ray energies were obtained independently from the previous measurements of muonic X-rays. 
The measured energies are consistent with the previous results of $^{208}$Pb.
As the muonic X-rays from $^{208}$Pb were accurately measured previously~\cite{Bergem1988-nf}, the calibration method conducted in this study confirmed the energy range to be below 6 MeV.
The K, L, M, and N lines were measured from $^{197}$Au target and the muonic K, L, M, N, and O lines were measured from $^{209}$Bi target.
All the energies of muonic X-rays obtained in this study are also consistent with those reported in the previous studies.

For the intensities, when comparing the relative values reported in this study and those reported by Measday et al.~\cite{Measday2007-zh}, the L, M, and N lines are consistent with the previous values, except for $6f_{7/2}$-$3d_{5/2}$ and $4d_{3/2}$-$2p_{1/2}$ transitions; however, a large discrepancy was found in the intensities of K rays in $^{197}$Au. 
This discrepancy is attributed to the overestimated efficiency curve in the previous measurement in high-energy regions. 
The intensity results of the present study are consistent with those reported by Hartmann et al.~\cite{Hartmann1982-wi}, except for $2p_{1/2}$-$1s_{1/2}$ and $5g$-$4f$ transitions. The energy of $2p_{1/2}$-$1s_{1/2}$ was 5.6 MeV, and the discrepancy may be attributed to the uncertainty of the efficiency of the detector. The energy of $5g$-$4f$ was approximately 400 keV, and the effect of self-absorption was significant. 

For $^{209}$Bi, the discrepancies in the intensities of the K and L lines, and the transitions in 449 keV and 823 keV were found between the present and previous studies. 
Measday et. al.~\cite{Measday2007-zh} reported the normalized intensities with some assumptions of the probability of a nonradiative transition and a nuclear-exciting transition~\cite{Measday2007-zh} for the total probability of the K and L transition series. The discrepancy may be attributed to the normalization assumptions in the previous research. 
For the $4d_{3/2}$-$2p_{1/2}$ transition, the previous research may have included the intensity of another peak at approximately 3696 keV. 
Therefore, the values reported in the present study are more reliable in terms of relative intensities because no assumptions were made.

Hence, we infer that the results of energies and relative intensities of muonic X-rays from $^{197}$Au and $^{209}$Bi provide new calibration reference values for photon detectors in high-energy regions below 8 MeV. 
The uncertainty of the result of this work provided in Table 4,5,6 includes systematic uncertainty in the energy column and excludes systematic uncertainty in the intensity column to be used practically for the calibration. The breakdowns of the uncertainties are written in the footnote of each Table.

\begin{table}[width=0.5\textwidth, cols=4,pos=h]
  \caption{Muonic X-ray energies for $^{208}$Pb.}
  \label{tab:muX_Pb}
  \begin{tabular*}{\tblwidth}{@{} LLLL@{} }
    \toprule
    & & Energy (keV) & \\
    transition & this work$^*$ & Kessler\cite{Kessler1975-sw} & Bergem\cite{Bergem1988-nf}\\
    \midrule
    $2p_{1/2}$-$1s_{1/2}$ & 5777.44(51) & 5777.91(40)   & 5778.058(100) \\
    $2p_{3/2}$-$1s_{1/2}$ & 5962.68(41) & 5962.77(42)   & 5962.854(90)  \\
    $3d_{3/2}$-$2p_{3/2}$ & 2457.72(46) & 2457.20(20)   & 2457.569(70)  \\
    $3d_{5/2}$-$2p_{3/2}$ & 2500.49(34)  & 2500.33(6)    & 2500.590(30)  \\
    $3d_{3/2}$-$2p_{1/2}$ & 2642.13(35)  & 2642.11(6)    & 2642.332(30)  \\
    $3p_{1/2}$-$2s_{1/2}$ & 1460.84(44) &      -        & 1460.558(32)  \\
    $3p_{3/2}$-$2s_{1/2}$ & 1507.44(37)  & 1507.48(26)   & 1507.754(50)  \\
    $4f_{5/2}$-$3d_{5/2}$ & -$^{a}$     & -             & 928.883(14)   \\
    $4f_{7/2}$-$3d_{5/2}$ & 938.04(32)   & 937.98(6)     & 938.096(18)   \\
    $4f_{5/2}$-$3d_{3/2}$ & 972.00(32) & 971.85(6)     & 971.974(17)   \\
    $5f_{7/2}$-$3d_{5/2}$ & 1366.55(38)$^{b}$  & 1366.52(8)    & 1366.347(19)  \\
    $5f_{5/2}$-$3d_{3/2}$ & 1404.94(46) & 1404.74(8)    & 1404.658(20)  \\
    $4d_{3/2}$-$3p_{3/2}$ & 874.09(61)&      -        & 873.761(63)   \\
    $4d_{5/2}$-$3p_{3/2}$ & -           &      -        & 891.383(22)   \\
    $4d_{3/2}$-$3p_{1/2}$ & -           &      -        & 920.959(28)   \\
    \bottomrule
  \end{tabular*}
  \leftline{*the 0.3-keV systematic uncertainty is included.}
  \leftline{$^{a}$The energy peak at 928 keV is masked by the tail of 938 keV.}
  \leftline{$^{b}$The energy peak at 1366 keV is deduced from the spectrum}
  \leftline{~obtained at 100 ns from the muon irradiation time because it is}
  \leftline{~overlapped with a 1368-keV delayed $\gamma$-ray from $^{24}$Na created}
  \leftline{~by muon capture reaction in the target holder ($^{27}$Al).}
\end{table}

\begin{table*}[width=0.9\textwidth, cols=4,pos=h]
  \caption{Muonic K, L, M, N series energies and intensities for $^{197}$Au.}
  \label{tab:muX_Au}
  \begin{tabular*}{\tblwidth}{@{} LLLLLLL@{} }
    \toprule
    & \multicolumn{3}{c}{Energy (keV)} & \multicolumn{3}{c}{Relative Intensity **}\\
    transition & this work* & Acker\cite{Acker1966-kh} and others & Measday\cite{Measday2007-zh} & 
    this work & Measday\cite{Measday2007-zh} & Hartmann~\cite{Hartmann1982-wi}\\
    \midrule
    $2p_{1/2}$-$1s_{1/2}$ & 5591.0(11)     & 5592.8(50)      & (5591.0)$^e$ & 86.2(10)  & 71.4(63)  & 76.1(42)\\
    $2p_{3/2}$-$1s_{1/2}$ & 5764.7(11)     & \multirow{2}{*}{5762.5(50)} & (5763.1)$^e$ & 89.6(15) & \multirow{2}{*}{115.2(84)} & \multirow{2}{*}{127.6(74)}\\
                          & 5746.3(11)     &                             &              & 38.68(94) & & \\
    $3p_{1/2}$-$1s_{1/2}$ & -              &      -          & \{8085\}$^f$ & -         & 3.3(33)   & \multirow{2}{*}{4.39(55)}\\
    $3p_{3/2}$-$1s_{1/2}$ & 8130.4(37)     &      -          & \{8128\}$^f$ & 1.78(59)  & 8.1(42)   & \\
    $3d_{3/2}$-$1s_{1/2}$ & 8065.3(34)     & -               & -            & 2.63(67)  & -         & \multirow{2}{*}{5.92(66)}\\
    $3d_{5/2}$-$1s_{1/2}$ & 8103.2(31)     & -               & -            & 3.91(72)  & -         & \\
    $3d_{3/2}$-$2p_{3/2}$ & 2304.25(68)    &    -            & 2302(2)      & 6.20(39)  & 8.6(35)   & \\
                          & 2319.86(66)    &                 &              & 7.36(51)  & & \\
    $3d_{5/2}$-$2p_{3/2}$ & 2341.52(57)    & \multirow{2}{*}{2343.1(25)} & 2341.2(2)  & 63.3(13) & \multirow{2}{*}{95.6(73)}& \\
                          & 2357.69(58)    &                             &            & 30.60(59) & & \\
    $3d_{3/2}$-$2p_{1/2}$ & 2477.27(59)    & 2474.4(20)      & (2477.8)$^e$ & 66.39(57)   & 63.3(75)  & $^{sum}$179.8(60)\\
    $4d_{5/2}$-$2p_{3/2}$ & 3203.61(90)    &    -            & 3202(5)      & 3.15(86)  & \multirow{2}{*}{6.9(21)} & \\
                          & 3221.4(12)     &                 &              & 1.28(64)  & & \\
    $4d_{3/2}$-$2p_{1/2}$ & 3361.10(97)    &   -             & 3356(5)      & 3.6(14)   & 7.7(25)   & $^{sum}$9.0(36)\\
    $5d_{5/2}$-$2p_{3/2}$ & -              &   -             & \{3601\}$^f$ & -         & 2.7(27)   & -\\
    $5d_{3/2}$-$2p_{1/2}$ & -              &   -             & \{3762\}$^f$ & -         & 2.1(21)   & -\\
    $3p_{1/2}$-$2s_{1/2}$ & 1391.31(72)    & 1392.31(48)$^c$ &     -        & 0.87(21)  & -         & \multirow{2}{*}{3.9(11)}\\
    $3p_{3/2}$-$2s_{1/2}$ & 1435.72(82)    & 1436.95(37)$^c$ &     -        & 3.7(16)   & -         & \\
    $4f_{5/2}$-$3d_{5/2}$ & 862.60(44)     & -               & -            & 6.28(40)  & -         & -\\
    $4f_{7/2}$-$3d_{5/2}$ & 870.04(31)     & 869.1(16)       & 870.11(10)   & 100.00(41)  & 100.0(23) & \multirow{2}{*}{172.6(35)} \\
    $4f_{5/2}$-$3d_{3/2}$ & 899.24(31)     & 899.6(14)       & 899.27(10)   & 72.65(41)   & 72.4(25)  &  \\
    $5f_{7/2}$-$3d_{5/2}$ & 1267.77(36)    &    -            & 1267(1)      & 12.38(40)   & 9.2(27)   & \multirow{2}{*}{18.21(78)}\\
    $5f_{5/2}$-$3d_{3/2}$ & 1300.70(40)$^a$&    -            & 1299(1)      & -$^a$     & 3.8(17)   &  \\
    $6f_{7/2}$-$3d_{5/2}$ & 1483.77(64)    &    -            & \{1482\}$^f$ & 2.84(60)  & 0.84(84)  &  \multirow{2}{*}{7.49(69)}\\
    $6f_{5/2}$-$3d_{3/2}$ & -              &    -            & \{1516\}$^f$ & -         & 0.84(84)  & \\
    $7f_{7/2}$-$3d_{5/2}$ & -              &    -            & \{1612\}$^f$ & -         & 1.7(10)   &  \multirow{2}{*}{3.02(73)}\\
    $7f_{5/2}$-$3d_{3/2}$ & -              &    -            & \{1647\}$^f$ & -         & 0.84(84)  & \\
    $5g_{9/2}$-$4f_{7/2}$ & 400.03(30)     &400.14(5)$^d$    & 400.15(15)   & 76.13(52)   & 81.0(44)  &  \multirow{2}{*}{154.4(57)}\\
    $5g_{7/2}$-$4f_{5/2}$ & 405.52(31)     &405.65(5)$^d$    & 405.58(15)   & 63.53(27)   & 63.9(44)  & \\
    $6g_{9/2}$-$4f_{7/2}$ & 615.31(33)     &     -           & 615.5(4)     & 11.30(66)   & 14.4(42)  &  \multirow{2}{*}{20.90(96)}\\
    $6g_{7/2}$-$4f_{5/2}$ & 621.80(33)     &     -           & 621.7(4)     & 10.86(42)   & 12.1(42)  & \\
    $7g_{9/2}$-$4f_{7/2}$ & 745.12(36)     &     -           & 744.9(5)     & 2.56(25)  & 6.1(61)   &  \multirow{2}{*}{5.85(23)}\\
    $7g_{7/2}$-$4f_{5/2}$ & 752.19(37)     &     -           & 752.1(5)     & 2.44(34)  & 4.8(23)   & \\
    $8g_{9/2}$-$4f_{7/2}$ & 828.38(58)$^b$ &     -           & \{829\}$^f$  & -$^b$     & 1.25(63)  &  \multirow{2}{*}{2.42(85)}\\
    $8g_{7/2}$-$4f_{5/2}$ & 836.09(47)$^b$ &     -           & \{836\}$^f$  & -$^b$     & 2.7(13)   & \\
    \bottomrule
  \end{tabular*}
  \leftline{~* the 0.3-keV systematic uncertainties are included in the energy region below 1.3 MeV and the energy depend uncertainties are included} 
  \leftline{~~~in the energy region above 1.3 MeV, for example, 1.0 keV systematic uncertainty is included in the $2p_{1/2}$-$1s_{1/2}$ transition.}
  \leftline{~* the listed energies are a weighted average of hyperfine splitting. These energy peaks show a larger FWHM}
  \leftline{~~~than the intrinsic energy resolution of the Ge detector.}
  \leftline{~**the intensity is normalized at 100\% at the transition of $4f_{7/2}-3d_{5/2}$ which shows the largest intensity.}
  \leftline{~**the intensities reported by Hartmann et al. are renormalized at the total intensity of $4f_{5/2}-3d_{3/2}$ and $4f_{7/2}-3d_{5/2}$}
  \leftline{~~~to 172.6\%, which is the relative intensity value reported in this study for the ease of comparison.}
  \leftline{~**multiply with 0.44 (4) to obtain the absolute intensity (See text).}
  \leftline{~blank energies were not detected because of low statistics.}
  \leftline{~$^a$1300 keV was overlapped with a DE peak of $3d_{5/2}$-$2p_{3/2}$ transition (1297.9 keV)}
  \leftline{~$^b$828 and 836 keV peaks are difficult to deduce intensities because of low statistics.}
  \leftline{~$^c$Powers(1974)~\cite{Powers1974-aa}}
  \leftline{~$^d$Engfer(1974)~\cite{Engfer1974-km}}
  \leftline{~$^e$ $K_\alpha$ lines and 2477.8 keV are used as energy calibration references, reported by Measday et al~\cite{Measday2007-zh}.}
  \leftline{~$^f$ curly brackets are the estimated values obtained from calculations.}
\end{table*}

\begin{table*}[width=0.8\textwidth, cols=4,pos=h]
  \caption{Muonic K, L, M, N, O series energies and intensities for $^{209}$Bi.}
  \label{tab:muX_Bi}
  \begin{tabular*}{\tblwidth}{@{} LLLLLL@{} }
    \toprule
    & \multicolumn{3}{c}{Energy (keV)} & \multicolumn{2}{c}{Relative Intensity **} \\
    transition & this work* & Engfer\cite{Engfer1974-km} and others & Measday\cite{Measday2007-zh} & this work & Measday\cite{Measday2007-zh} \\
    \midrule
    $2p_{1/2}$-$1s_{1/2}$ & 5841.48(35)  & 5839.7(55)$^c$ & 5841.5(30) & 98.51(97) & 84.5(35) \\%
    $2p_{3/2}$-$1s_{1/2}$ & 6028.25(37)  & \multirow{2}{*}{6032.2(50)$^c$} & \multirow{2}{*}{6032.4(30)} & 69.6(12) & \multirow{2}{*}{113.8(35)} \\
                          & 6038.83(36)  &                                 &                             & 70.88(89) & \\
    $3p_{3/2}$-$1s_{1/2}$ & 8629.2(27)   &                & 8628(10)   & 1.48(58)  & 1.64(94) \\
    $3d_{3/2}$-$1s_{1/2}$ & 8542.3(18)   &                & 8539(10)   & 2.68(62)  & 5.39(70) \\
    $3d_{5/2}$-$1s_{1/2}$ & 8587.2(17)   &                & 8584(10)   & 4.92(68)  & 6.09(70) \\
    $3d_{3/2}$-$2p_{3/2}$ & 2504.22(49)  & 2501.81(59)    & 2504(2)    & 7.3(11) & 3.5(19) \\
                          & 2513.57(50)  &                &            & 7.6(14) & \\
    $3d_{5/2}$-$2p_{3/2}$ & 2550.15(33)  & \multirow{2}{*}{2554.8(20)$^c$} & 2549.6(3)  & 50.12(72) & 52.7(26) \\
                          & 2559.50(34)  &                                 & 2558.9(3)  & 38.74(90) & 43.1(26) \\
    $3d_{3/2}$-$2p_{1/2}$ & 2700.30(32)  & 2700.50(17)    & 2700.3(2)  & 66.88(44) & 62.5(26) \\
    $4d_{5/2}$-$2p_{3/2}$ & -$^a$        &      -         & 3510(1)    & -         & 9.4(33) \\
    $4d_{3/2}$-$2p_{1/2}$ & 3678.78(61)  &       -        & 3679(1)    & 2.64(31)  & 5.2(14) \\
    $5d_{5/2}$-$2p_{3/2}$ &  -           &        -       & \{3951\}   & -         & 0.7(14) \\
    $5d_{3/2}$-$2p_{1/2}$ &  -           &         -      & \{4132\}   & -         & 1.2(12) \\
    $4f_{7/2}$-$3d_{5/2}$ & 961.28(31)   &  961.18(25)    & 961.3(2)   & 100.00(35)& 100.0(68) \\
    $4f_{5/2}$-$3d_{3/2}$ & 996.76(31)   &  996.67(25)    & 996.8(2)   & 79.31(35) & 80.8(56) \\
    $5f_{7/2}$-$3d_{5/2}$ & 1400.49(35)  &        -       & 1400.4(5)  & 9.34(29)  & 9.1(14) \\
    $5f_{5/2}$-$3d_{3/2}$ & 1440.33(41)  &        -       & 1440.2(5)  & -$^d$     & 6.8(19) \\
    $6f_{7/2}$-$3d_{5/2}$ & 1639.97(39)  &        -       & 1640.1(5)  & 4.07(20)  & 5.2(14) \\
    $6f_{5/2}$-$3d_{3/2}$ &  -$^b$       &        -       & \{1681.0\} & -         & 2.8(23) \\
    $7f_{7/2}$-$3d_{5/2}$ &  -           &       -        & 1783(2)    & -         & 0.94(47) \\
    $7f_{5/2}$-$3d_{3/2}$ &  -           &       -        & 1826(2)    & -         & 1.9(14) \\
    $8f_{7/2}$-$3d_{5/2}$ &  1876.46(50) &       -        & 1876(2)    & 1.34(16)  & 2.3(14) \\
    $8f_{5/2}$-$3d_{3/2}$ &  -           &      -         & 1920(2)    & -         & 1.6(16) \\
    $5g_{9/2}$-$4f_{7/2}$ & 442.09(30)   &  442.107(50)   & 442.0(1)   & 95.18(86) & 93.2(44) \\
    $5g_{7/2}$-$4f_{5/2}$ & 448.76(30)   &  448.828(50)   & 448.7(1)   & 80.5(10)& 74.9(44) \\
    $6g_{9/2}$-$4f_{7/2}$ & 679.75(32)   &         -      & 679.7(1)   & 10.75(22) & 11.2(23) \\
    $6g_{7/2}$-$4f_{5/2}$ & 687.55(33)   &         -      & 687.6(2)   & 7.61(19)  & 8.7(23) \\
    $7g_{9/2}$-$4f_{7/2}$ & 823.38(37)   &         -      & 823.4(5)   & 5.52(40)  & 3.3(12) \\
    $7g_{7/2}$-$4f_{5/3}$ & 831.77(39)   &        -       & 832.0(5)   & -$^d$     & 2.8(14) \\
    $6h$-$5g$             & 239.60(30)   &  239.1(16)$^c$ &    -       & 179.08(52)& - \\
    \bottomrule
  \end{tabular*}
  \leftline{~* the 0.3-keV systematic uncertainty is included the energy region lower than 8 MeV, and the 1.0-keV uncertainty is included}
  \leftline{~~~in the energy region above 8 MeV.}
  \leftline{~* the listed energies are a weighted average of hyperfine splitting. These energy peaks show a larger FWHM}
  \leftline{~~than the intrinsic energy resolution of the Ge detector.}
  \leftline{~**the intensity is normalized at 100\% at the $4f_{7/2}-3d_{5/2}$ transition, which shows the largest intensity at 1 MeV.}
  \leftline{~**multiply with 0.42 (4) to obtain the absolute intensity (See text).}
  \leftline{~$^a$Peaks around 3510 keV were difficult to distinguish from other peaks.}
  \leftline{~$^b$$6f_{5/2}$-$3d_{3/2}$ transition energy peak was overlapped by the double escape peak of $3d_{3/2}$-$1p_{1/2}$ transition. }
  \leftline{~Other blank peaks were not detected because of the low intensity.}
  \leftline{~$^c$Acker(1966)~\cite{Acker1966-kh}}
  \leftline{~$^d$The intensities of 1440 and 832 keV were not shown because other peaks were contained in the tail, and uncertainties are large.}
\end{table*}

\section{Summary} \label{sec:summary}
In this study, we have developed a wide-energy-range photon detection system for muonic X-ray spectroscopy.
The detection system includes photon detectors consisting of high-purity Ge detectors and BGO Compton suppressors, a data acquisition system based on the waveform digitizer. 
A calibration method for the photon detector are also provided.

Three types of Ge detectors were used to cover a wide energy range.
The detector performance of the proposed system in the entire energy range below 10 MeV was evaluated through offline source measurements and in-beam experiments using the ${}^{27}\mathrm{Al}(p,\gamma)^{28}\mathrm{Si}$ resonance reaction.
Optimization of the parameter used in the digital waveform processing was essential to achieve the best performance of the system.
With a sufficient number of anchor points for energy calibration, an energy accuracy of 0.3 keV was achieved.
By selecting the appropriate timing pick-off method, a timing resolution ($\sigma$) of 10--20 ns was obtained.
The absolute efficiency was also determined with 3\% accuracy.
The detector performances under high count-rate conditions of up to 10~kHz were investigated.
The Monte-Carlo based simulation using the GEANT4 was performed and the obtained results were in agreement with the measurement results. 
This indicates that configurations of each spectroscopy setup can be optimized using GEANT4.

The performance of the system was demonstrated at PSI.
The performance of BGO Compton suppressors was evaluated through meteorite measurements conducted at PSI by improving the S/N ratio by a factor of 1.9 and 3.9, depending on the energy region. 
The carbon component of the Allende meteorite with 0.3~wt\% could be detected. 
We proposed a calibration method for the photon detector at the muon facility using muonic X-rays emitted from Au and Bi.
The energy and intensity references proposed in this study provide a method for performing general calibrations at the muon facility.


\section*{Acknowledgements}
This work was supported by the Japan Society for the Promotion of Science (JSPS) KAKENHI Grant Number 18H03739, 19H01357, 22H02107, 22K18735 and Swiss National Science Foundation Synergia project "Deep$\mu$" Grant Number 193691. R. M. is supported by the Fore-front Physics and Mathematics Program to Drive Transformation (FoPM), a World-leading Innovative Graduate Study (WINGS) Program, and the JSR Fellowship from the University of Tokyo. 
This work was partly supported by the RCNP Collaboration Research Network program with project number COREnet-41.
We thank the SUNFLOWER collaboration for providing the Au and Bi targets. 
A part of the experiment was performed at the Pelletron facility (joint-use equipment) at the Wako Campus, RIKEN.











\bibliographystyle{elsarticle-num}

\bibliography{References.bib,reference_manual.bib}

\bio{}
\endbio

\endbio

\end{document}